\documentclass[12pt]{article}
\usepackage[multiple]{footmisc}
\usepackage[margin=1in]{geometry}
\usepackage{setspace}
\usepackage{amssymb, amsmath, amsthm}
\usepackage{natbib}

\newtheorem{assumption}{Assumption}
\newtheorem{example}{Example}

\newtheorem{observation}{Observation}

\newtheorem{proposition}{Proposition}
\newtheorem{corollary}{Corollary}

\usepackage[hidelinks]{hyperref}

\let\citeasnoun=\citet

\begin{document}

\sloppy 

\title{Parallel Trends and Dynamic Choices%
\thanks{
We thank the editor and three referees for their helpful comments.
We also thank Aureo de Paula, Isma\"el Mourifi\'e, Jonathan Roth, Pedro Sant'Anna and participants at the 2023 Northwestern University Econometrics Day and the 2022 SEA Conference.
All errors are our own.
Edited by Christian Hansen.
}
}
\author{ Philip Marx\footnote{Louisiana State
    University. Email: \href{mailto:pmarx@lsu.edu}{pmarx@lsu.edu}.}
\and  Elie Tamer\footnote{Harvard University. Email:
    \href{mailto:elietamer@fas.harvard.edu}{elietamer@fas.harvard.edu}.}
\and Xun Tang\footnote{Rice University.
    Email: \href{mailto:xt9@rice.edu}{xt9@rice.edu}.}
}

\begin{spacing}{1}
\maketitle

\begin{abstract}
Difference-in-differences is a common method for estimating treatment effects,
and the parallel trends condition is its main identifying assumption: the trend in mean untreated
outcomes is independent of the observed treatment status. In observational settings,
treatment is often a dynamic choice made or influenced by rational actors, such as
policy-makers, firms, or individual agents. This paper relates parallel trends to
economic models of dynamic choice. We clarify the implications of parallel trends on
agent behavior and study when dynamic selection motives lead to violations of parallel
trends. Finally, we consider identification under alternative assumptions that
accommodate features of dynamic choice.
\end{abstract}

\vspace{0.25in}
\begin{quote}
\end{quote}

\end{spacing}

\newpage 
\section{Introduction}
Difference-in-differences (or {\it DiD}) is a popular research design for causal inference with panel data or repeated cross-section data. Though econometricians have estimated panel data regressions for decades,  the  causal interpretation of coefficients from these regressions is subtle and forms the basis of this paper.  On the one hand, and with non-experimental data, structural choice models with  optimizing agents provide causal interpretations.
These depend on what one assumes about how these choices (about treatment, inputs, or prices) were made. On the other hand, and without a behavioral model, identifying assumptions are used to provide causal interpretations that allow for evaluations of various treatments.
Chief among them is an assumption of parallel trends, namely that the untreated trend is independent of the realized sequence of treatment.
Such assumptions indirectly restrict the kinds of choice behavior that agents are allowed to have. This is especially important in the dynamic settings where these data and designs are mainly used. 

In this paper, we clarify the connections between the central DiD causal identifying assumption, i.e., parallel trends, 
and dynamic choice behavior.
We do so mainly by highlighting through simple examples the kinds of choice behaviors that are or are not compatible with the assumption.
As a second contribution, we consider identification under alternative assumptions that accommodate various features of our discussion on dynamic choice. 

A complicating factor that raises various potential selection issues in DiD models is time. 
This is especially so when treatments are indexed by time in observational data. 
In that case, a variety of dynamic considerations, such as future discounting, option values, learning, and anticipation, may all play an important role in determining choices.
While these selection and information issues and their role in empirical models are well known to econometricians, the purpose of this paper again is to shed light on exactly how these behaviors intersect with commonly used assumptions in DiD models, as well as what classes of economic choice models of optimizing behavior are consistent with the behaviorally agnostic DiD model.

For example, we show that some dynamic features, such as discounting or time-varying costs of treatment, need not lead to violations of parallel trends per se. On the other hand, other phenomena such as learning, optimal stopping, or repeated Roy-model-like behavior can potentially violate parallel trends. 
We also elucidate how and why parallel trend assumptions among some subsequences of observed treatment may be more robust to dynamic selection concerns than others. 
Finally, we translate these observations into alternative identification assumptions and strategies. These include considering weakened parallel trend assumptions, time stationarity assumptions that can circumvent dynamic selection concerns, and partially identifying assumptions motivated by our structural models of choice.

We contribute primarily to a recent yet already substantial literature clarifying aspects of DiD assumptions and estimators in settings that extend the basic two-period DiD model with no treatment in the initial period.
Much of this work has focused on i) interpreting typical regression-based slope parameters and what these estimate when heterogeneity in effects is allowed over multiple time periods, and ii) deriving estimators that target a causal parameter of interest under parallel trends.%
\footnote{See for instance \cite{de2018fuzzy}, \cite{CD}, \cite{borusyak}, \cite{callawaysatanna}, \cite{goodman2021difference}, \cite{sundid}, and \cite{athey2018designdid} to name a few. With the exception of the first two papers, most of this work considers staggered designs where the sequence of realized treatment is monotonic across time.}  
One takeaway from this literature is that a variety of identification and estimation complications arise in the empirically common settings where the basic DiD model is enriched to fuzzy designs (where some individuals are treated in the ``pre-treatment'' period) or multiple time periods. 
Yet, it is precisely also in these richer settings that we might expect richer dynamic selection considerations to arise.
Thus, we complement this literature by focusing on the relationship between parallel trends and structural models of dynamic choice. 

Our exercise is closest in spirit to two recent papers. In concurrent work, \cite{ghanem2022selection} study the relationship between selection and the parallel trends assumption. 
They consider the basic design where no unit is treated initially and selection on untreated outcomes in the following period can be a function of an individual fixed effect or time-varying errors in the untreated outcome equation.
We extend the class of selection mechanisms under consideration in two ways. First, we consider selection in fuzzy designs, which allows us to further explore the dynamics of selection with two periods of treatment. 
In addition, we explicitly model selection under imperfect information, which allows us to consider models with a temporal resolution of uncertainty and learning. 
Finally, we consider alternative identification approaches in the case where the parallel trends assumption remains in question.

In other related work, \cite{fudenberg2022learningte} study how learning by decision-makers affects the identification and interpretation of treatment effects. 
They consider a model where decision-makers choose an amount of effort in each period that affects both the treatment and the outcome. Decision-makers learn about the returns to effort from treatment assignment, and outcomes depend on effort but not treatment per se. 
In their model, the DiD estimator identifies the sum of a preference effect and a learning effect but does not separate the two effects. 
In our learning example of \autoref{sec:learning}, we model treatment as the choice variable and past outcomes as potentially informative signals, and we investigate when learning does (not) lead to violations of the parallel trends assumption.

More broadly, our work fits into a literature that investigates and offers alternative identification methods for when parallel trends fails. 
Earlier investigations into modeling selection in panels include \cite{ashenfeltercard1985training} and \cite{heckmanrobb1985}.
Other work has studied the identified features of the model under weaker assumptions by deriving bounds on parameters of interest (\cite{manski2018rtc}, \cite{rothrambachan}, and \cite{bankedagni2022}). 
\cite{abadie2005xdid} considers identification under covariate-conditional parallel trends and offers related examples for when such a conditional approach may be more credible.
In the presence of pre-trends, \cite{dobkin2019hospital} consider extrapolations of pre-trends as an alternative to parallel trends, while \cite{freyaldenhoven2019pretrend} propose an alternative identification approach based on the existence of covariates that relate to the policy of interest only through the pre-trend confounder. 
Similar in spirit to our work, \cite{freyaldenhoven2021viz} also consider the economic content of various identifying assumptions in an event study framework and in the presence of an unobserved confounder.
Complementary to this work, our discussion of identification focuses on alternatives that do not require any additional data, such as pre-trends, proxies, or instrumental variables. In this sense, our approach is also similar yet complementary to the recent doubly-robust approach to identification in panel data models of \cite{arkhangelskyimbens2021doubly}. 

In addition, we draw on a large literature in applied economics that explicitly models dynamic decisions capturing incentives, option values, and/or learning. See for instance \citeasnoun{heckman1976life}, \citeasnoun{rust1994structural}, and \citeasnoun{keanewolpin}. These models are central to the literature on structural estimation of dynamic decision processes.\footnote{These models are directly related to the literature on reinforcement learning and causal bandits. See \citeasnoun{igamiAI} for more on this connection.} See also the work of \citeasnoun{taber}, \citeasnoun{heckmannavaro} and \citeasnoun{abbring2010identification}.  This structural literature embeds within it causal parameters because it builds a model of behavior that is based on economic optimizing agents. Hence, causal relations are provided by construction through model specification that can be used to address counterfactuals and policy impacts. The literature on DiD on the other hand focuses on identifying parameters that are given a causal interpretation under some identifying restrictions without the need for full specification of a behavioral model. We think this is a strength of the approach and one that is worth highlighting. However, we also point out that care should be taken in particular setups when these ``design'' assumptions come at odds with simple notions of dynamic selection that may be relevant in   data applications. 

Finally, the exercise that we undertake is familiar to econometricians\footnote{For a similar exercise on the role theory plays, more precisely general equilibrium, in formulating empirical models, see \cite{acemoglu2010theory}.}. 
In standard static causal setups, \cite{vytlacil} provided a class of choice models that are consistent with the LATE assumptions of \cite{imbensandangrist}. 
In particular, \cite{vytlacil} showed that a separable choice model with a univariate unobservable is (observationally) equivalent to the monotone causal model underlying LATE. 
A univariate unobservable determining choices (or monotonicity) may well be a reasonable model of behavior in many applications. Still, the existence of such a choice model clarifies the types of behavior under which causal interpretations are warranted. 
While our paper does not attempt a choice-theoretic characterization of the DiD model, it takes an otherwise similar approach in the context of dynamic panel/DiD regressions. 

The paper proceeds as follows. 
In \autoref{sec:model} we review the two-period fuzzy DiD model and the parallel trends assumption that the mean untreated trend is independent of the entire realized sequence of treatment. 
In \autoref{sec:simple_example} we begin by introducing a simple but useful equivalent formalization of the ``full'' parallel trends assumption as a set of conditional or ``partial'' assumptions on subsequences of treatment. 
In \autoref{sec:examples} we use this to study a sequence of choice models and their compatibility with parallel trends. 
In \autoref{sec:id} we then provide some other suggestive approaches that either weaken parallel trends or replace it with more selection-robust or selection-motivated assumptions.
The emphasis throughout is on stylized models that retain key features of dynamic selection in a simple framework that allow one to focus on exact channels of interest. 
\autoref{sec:conclusion} concludes. 
Additionally, \autoref{apx:proofs} collects the proofs, \autoref{apx:heuristics} provides a summary list of guidelines and heuristics that highlight when parallel trends is or is not plausible in an empirical setting,
and 
\autoref{apx:illustrate} discusses these various heuristics in an example based on \cite{ashenfeltercard1985training}.

\section{Model} 
\label{sec:model}

For each of $N$ individuals indexed by $i$ in a pair of time periods
indexed by $t \in \{0,1\}$, 
a researcher observes a sequence of realized
treatments $D_{it} \in \{0,1\}$ and realized outcomes $Y_{it} \in \mathbb{R}$.
We restrict our analysis to two time periods to clearly highlight linkages and
for ease of exposition. 
We assume potential outcomes for each individual are indexed by that individual's
treatment in the present period only, and individuals are randomly sampled from
a population. 
We index this assumption by zero since we maintain it throughout most of our
investigation of selection in treatment and its relationship to parallel trends.

\setcounter{assumption}{-1}
\begin{assumption}
    [Limited Dependence on Treatment and Random Sampling]
\label{assn:present}
Potential outcomes at time $t$ depend only on own treatment at time $t$, i.e.
$Y_{it}(\mathbf d_1, \ldots, \mathbf d_N ) = Y_{it}(d_{it})$ for $t \in \{0,1\}$ and $\mathbf d_i = (d_{i0}, d_{i1}),$ with $i = 1, \dots, N$.
Individuals are randomly sampled from a population model $(Y_0 (\cdot), Y_1(\cdot), \mathbf{U})$, where $Y_t (\cdot) \equiv (Y_t (0), Y_t (1))$ denotes the pair of potential outcomes in period $t$, and $\mathbf{U}$ denotes a random vector that contains
decision-relevant variables which may also be indexed by the time period or the sequence of treatment. 
\end{assumption}

In addition to imposing SUTVA, \autoref{assn:present} requires that potential outcomes in a period depend only on treatment in that period.
This rules out dynamic treatment effects and some forms of anticipation, but suffices for exploring the dynamics of selection into treatment.%
\footnote{
    Allowing for anticipation effects, whereby current potential outcomes can depend on future treatments, poses challenges to identification in dynamic choice models. This has received attention in the literature. 
    See for instance \cite{abbringvandeberg}, \cite{heckmannavaro} and \citeasnoun{heckman2016dynamic}, and in the context of DiD models, \cite{malani2015anticipation}.  
    We discuss how to relax the dependence on contemporaneous treatment alone in \autoref{sec:extra}.
}
We henceforth work in terms of the population variables and suppress the individual index $i$ to simplify notation.  

The central condition we investigate is that the means of the untreated outcomes, i.e., $Y_t(0)$'s, satisfy a parallel trends condition across each treatment sequence $(D_0,D_1)$. 
This is the key assumption maintained by the DiD literature. 

\bigskip 
\noindent 
\textbf{Parallel Trends (PT).} \textit{The mean untreated trend is constant across all realized sequences of treatment:}
    \[ 
    E [ Y_{1} (0) - Y_{0} (0) | D_0 = d_0, D_1 = d_1] = \tau 
    \]
\textit{for some constant} $\tau \in \mathbb{R}$ \textit{and all} $d_0, d_1 \in \{0,1\}$.
\bigskip 

Thus expressed, the parallel trends condition is identical to the \textit{strong exogeneity} condition for two time periods and a given group in \cite{CD}.
In the basic case with a pre-treatment period, a.k.a.~the \textit{sharp design}, where $D_0 = 0$ for all individuals, the parallel trends condition  only conditions on treatment in period 1. 
In that case and under this assumption, the causal effect of treatment on the treated in period 1 is identified by the basic DiD estimator:
\[
    E [ Y_{1} (1) - Y_{1} (0) | D_1 = 1 ]
    =
    E[ Y_{1} - Y_{0} | D_1 = 1 ] - E[ Y_{1} - Y_{0} | D_1 = 0] 
\] where $Y_t = D_tY_t(1) + (1-D_t)Y_t(0)$ is the realized outcome in period $t$.

A recent literature studies DiD estimators that identify weighted averages of heterogeneous treatment effects in  more general cases where some units are treated in period 0 (a.k.a. the \textit{fuzzy design}) or there are more than two periods.%
\footnote{
See for example, \cite{de2018fuzzy}, \cite{CD}, \cite{borusyak}, \cite{callawaysatanna}, \cite{goodman2021difference}, \cite{sundid}, and \cite{athey2018designdid}.}
Those papers impose versions of the parallel trends condition, in combination with further assumptions such as monotonicity of $D_t$ in $t$ (staggered research design) or some versions of parallel trends in treated outcomes. 

\section{Setup and a Simple Example}
\label{sec:simple_example}

Our first goal is to explore when the parallel trends condition is (in)consistent with treatments that are determined by forward-looking and rational dynamic choices.
In doing so, we focus on models that satisfy \autoref{assn:present}. 
This illustrates when individual motives leading to treatment selection are compatible with the parallel trends condition. 
We discuss generalizations in \autoref{sec:extra}. 

To start, notice that the parallel trends condition can be equivalently expressed as a set of pairwise equalities, which will be useful in analyzing subsequent examples.

\begin{observation}
    \label{thm:pt-dc}
    The parallel trends condition is equivalent to the two joint statements below:
    \begin{align}
        E [ Y_1 (0) - Y_0 (0) \mid D_0 = d_0] & \quad \text{is constant in $d_0$;} \label{eq:pt-dc1} \\[0.05in]
        E [ Y_1 (0) - Y_0 (0) \mid D_0 = d_0, D_1 = d_1] &\quad \text{is constant in $d_1$ for each $d_0$.}  \label{eq:pt-dc2}
    \end{align}
\end{observation}

\noindent
By construction, the two constants mentioned in (\ref{eq:pt-dc1}) and (\ref{eq:pt-dc2}) must be the same. In turn, each of these two equalities can be rearranged and reinterpreted as a stationarity condition in the magnitude of selection.

\begin{observation}
    \label{thm:pt-selection}
    Conditions \eqref{eq:pt-dc1} and \eqref{eq:pt-dc2} above are respectively equivalent to:
    \begin{align}
        E [ Y_t (0) \mid D_0 = 1] - E[ Y_t (0) \mid D_0 = 0] & \quad \text{is constant in $t$;} \label{eq:ss-dc1} \\
        E [ Y_t (0) \mid D_0 = d_0, D_1 = 1] - E [ Y_t (0) \mid D_0 = d_0, D_1 = 0] & \quad \text{is constant in $t$ for each $d_0$.} \label{eq:ss-dc2}
    \end{align}
\end{observation}

\noindent
Conditions \eqref{eq:ss-dc1} and \eqref{eq:ss-dc2} clarify how the parallel trends condition allows for flexible selection on the untreated outcome in a given period. 
This is evident because the two constants mentioned in \eqref{eq:ss-dc1} and \eqref{eq:ss-dc2} respectively are allowed to be different.
At the same time, the conditions require the magnitude of this selection (in terms of the mean untreated potential outcome) to be fixed across periods $t$ given a treatment history. 

It will also be instructive to illustrate parallel trends in a parametric setting where potential outcomes, $Y_{it}(d)$ with individual subscript $i$, are determined as in a panel data model: 
\begin{equation}
    \label{eq:outcome-model}
    Y_{it} (d) = \alpha_i + \delta_t + \beta_{it}d +\varepsilon_{it} \quad \text{ for } d \in \{0,1\}. 
\end{equation}
Then the parallel trends condition holds if:
\begin{equation}
\label{eq:demit}
    E [ \delta_1 - \delta_0 + \varepsilon_{i1} - \varepsilon_{i0} | D_{i0}=d_0, D_{i1}=d_1] 
    \quad \text{is constant in $(d_0, d_1)$}. 
\end{equation}
Thus, parallel trends holds if the evolution of time trends $\delta_1-\delta_0$ and the difference between transitory errors $\varepsilon_{i1} - \varepsilon_{i0}$ are mean-independent of observed treatments. 
This shows that parallel trends allow for some correlation between transitory errors and observed treatments, and hence allow for some selection on treatment, as long as (\ref{eq:demit}) holds.%
\footnote{
    \cite{abadie2005xdid} makes a similar observation for the case with sharp designs when $D_{i0} = 0$.}

For the rest of this section and \autoref{sec:examples}, we investigate the limits of endogenous selection into treatment allowed under parallel trends. 
We begin by revisiting a special case of selection on past outcomes.
In a classic example, \cite{ashenfelter1978} finds that participants in job training programs typically experienced a decline in earnings in the year prior to training.
As he observes, ``In retrospect this is not very surprising since the Department of Labor was instructed to enroll unemployed workers in the MDTA [training] programs in this period and it is just such workers who would be most likely to want to enter a training program.''

The example below presents a stylized version  for models of this phenomenon in \cite{ashenfeltercard1985training} and \cite{abadie2005xdid}; 
we focus on a two-period case where no one is treated in period 0 (i.e., there is a \textit{pre-treatment} period), potential outcomes are binary, and there is perfect selection on the past (untreated) outcome in period 1.
With a pre-treatment period, the parallel trends condition reduces to the familiar condition that $E[ Y_1 (0) - Y_0 (0) | D_1 = d_1 ]$ is constant in $d_1$.  
In this example, the parallel trends condition fails, except for the extreme case where the potential outcome without treatment does not vary over time. 

\begin{example}[Selection on Past Outcomes]
    \label{ex:violate-bs}
Potential outcomes are binary, i.e., $Y_t (0), Y_t(1) \in \{ 0, 1 \}$.
At $t=0$, everyone is untreated, i.e., $D_0=0$. 
At $t=1$, there is perfect selection on the previous period's realized outcome,
i.e., $D_1 = 1-Y_0$.
Assume $P(Y_0(0) = 0) \in (0,1)$.
Then the parallel trends (PT) condition holds if and only if the potential outcome with no treatment does not vary over time:
\[
    Y_0 (0) = Y_1 (0) \quad \text{a.s.}
\]
\end{example}

We present proofs of all claims in the examples in \autoref{apx:proofs}. 
Because the equality of untreated potential outcomes over time is a strong and often falsifiable condition, this simple example confirms that parallel trends are likely to fail when there is backward-looking dynamic selection: 
the propensity to enroll in a training program in the current period depends on employment in the previous period.
More generally, when current decisions depend directly (or indirectly) on past
outcomes, parallel trends are likely to fail.

In the next section, we provide an economic framework that is helpful in framing the parallel trends condition within a dynamic discrete choice model that is familiar to economists. 

\section{Parallel Trends under Dynamic Treatment Choices} \label{sec:examples}

We now introduce a dynamic model where treatment decisions are made by rational, forward-looking individuals who maximize the present value of expected utility over time.
This provides a unifying framework for the remaining examples of the section. 

To define the decision makers' maximization problem, we introduce two components. 
First, preferences over treatment histories $\mathbf d^t \equiv (d_0, \dots, d_t)$ are summarized by a flow of utilities $V_t(\mathbf d^t)$ in each period $t$, whose net present value is computed using a discount factor $\beta \in (0,1)$ that may vary across individuals.
In turn, per-period utility $V_t (\mathbf d^t)$ is separable into the potential outcome $Y_t (d_t)$ in that period, and an additional preference element $ K_t (\mathbf d^t)$ capturing the possibly history-dependent costs of (no) treatment. 
The costs $K \equiv (K_t (\mathbf d^t))_{t, d_0,d_1}$ are allowed to vary with potential treatment history, which accommodate staggered designs or switching costs. 
Second, the information available to the decision-maker consists of $\mathcal{I}_0$ in period 0 and $\mathcal{I}_1 (d_0)$ in period 1; the latter is allowed to depend on past treatment $d_0 \in \{0,1\}$.%

We assume $\mathcal{I}_0 \subseteq \mathcal{I}_1 (d_0)$. That is, the information available to the decision-maker accrues over time.
For simplicity, let the preference components be known to the decision-maker initially, i.e., $\beta, K \in \mathcal{I}_0$.

\bigskip
    
\noindent \textbf{Dynamic Utility Maximization (DUM).}
\textit{Given initial information} $\mathcal{I}_0$ \textit{at the start of period 0 and accrued information} $\mathcal{I}_1(d_0)$ \textit{at the start of period 1, the treatment decisions in periods 1 and 0 maximize the sum of expected discounted utility:}
\begin{align}
    D_1 (d_0) &= \mathbf{1} \{ E [ V_1 (d_0, 1) - V_1 (d_0, 0) | \mathcal{I}_1 (d_0) ] \geq 0 \}, \label{eq:opt_D1} \\
    D_0 &= \mathbf{1} \{ E [ V_0 (1) - V_0 (0) + \beta ( W_1 (1) - W_1 (0)) | \mathcal{I}_0 ] \geq 0 \},  \label{eq:opt_D0}
\end{align}
where $\mathbf{1} \{ \cdot \}$ denotes the indicator function and:
\begin{equation}
    W_1 (d_0) \equiv \max_{d_1 \in \{0,1\}} \, E [ V_1 (d_0, d_1) | \mathcal{I}_1 (d_0)], \label{eq:cont-val} 
\end{equation}
\textit{with} $V_t(\cdot)$ \textit{being the per-period utility, which is equal to a deterministic function of the realized outcome net a individual-varying cost of treatment:}%
\footnote{
    It is straightforward to generalize our results by letting the utility from outcomes using a function $f_t (Y_t (d_t))$; we refrain from doing so in order to simplify notation.
}
\begin{equation}
    \label{eq:utility-sep}
    V_t (\mathbf d^t) = Y_t(d_t) - K_t (\mathbf d^t). 
\end{equation}
\textit{Preference components are known to the decision-makers initially, i.e.,} $\beta, K_t(\cdot) \in \mathcal{I}_0$.

\bigskip

To present our analysis, it is helpful to write $\mathcal{I}_0 = \{ U_0 \}$ and $\mathcal{I}_1 (d_0) = \{ U_0, U_1 (d_0) \}$. 
Thus, $U_0$ denotes the initial information, and $U_1 (d_0)$ the incremental information accrued before period 1, which may be specific to past treatment $d_0$. 
In the notation of \autoref{assn:present}, this means $\mathbf{U}\equiv (U_0,U_1(0),U_1(1))$.
Furthermore, we let $U_1(d_0) = Y_0(d_0)$ in our examples below.
This is only for the purpose of simplifying exposition; our results extend readily to the cases where $Y_0(d_0)$ is a sub-vector of $U_1(d_0)$.

The econometrician observes the realized sequence of outcomes $(D_0, D_1, Y_0 , Y_1 )$.
Thus, the costs $K$ can also capture unobserved states in dynamic discrete choice models, e.g.~\cite{rust1994structural}.
The dependence of information on counterfactual treatment histories allows for information acquisition motives.  
We assume the decision-makers rationally form and update beliefs given knowledge of the joint distribution of unobservables and potential outcomes.

Consider two applications that fit in this setup.
The first is the entry and exit of firms. For example, \cite{gentzkow2011entry} use a first-difference approach, with state and year fixed effects, to estimate the effect of newspaper entry and exit on voter turnout in local markets.
The second is the effect of union membership on wages. 
This is commonly estimated in panel data by regressing worker wages on union membership, with worker and time fixed effects used to control for unobserved differences in worker productivity, e.g., \cite{jakubson1991unions}. 
In this case, workers may decide to join a union if the union wage premium exceeds costs of union membership, and the rational, forward-looking workers may take into account uncertainties about potential outcomes, particularly those from untested options.%
\footnote{
     Another example is \cite{vellaverbeek1998unions}, who estimated a dynamic Roy-type model, where workers locate each period in their preferred sector.}

Next, we investigate how the parallel trends (PT) condition can be consistent with such endogenous selection of treatment in the dynamic utility maximization (DUM) context.

\subsection{Sufficient Initial Information}
\label{sec:persist-sep}

We begin with a benchmark case of this model which satisfies the parallel trends condition.
In the subsequent section, we show how relaxing certain attributes of this benchmark case leads to violations of the parallel trends condition. 

For several examples in this and subsequent sections, we will maintain that the trends in the untreated potential outcomes is mean-independent from the initial information:
\begin{equation}
\label{eq:info-persist-pt}
      E [Y_1 (0) - Y_0 (0) \mid U_0] = \tau \text{ for some constant } \tau \in \mathbb{R}.
\end{equation}
\noindent Without this condition, one could expect the parallel trends condition to fail. For example, if the treatment selection is \textit{static} (e.g., when $D_0,D_1$ are determined by the initial information $U_0$ alone), then the parallel trends condition can fail as $E[Y_1(0)-Y_0(0)\mid U_0]$ varies over the support of $U_0$ partitioned by realizations of $(D_0,D_1)$.
Nonetheless, our goal is to investigate the parallel trends condition under \textit{dynamic} endogenous selection (when rational forward-looking individuals self-select into treatments).
Therefore, we maintain \eqref{eq:info-persist-pt} in several examples below so that the role of these dynamic motives can be isolated and manifest.

\begin{example}[No Learning about Potential Outcomes]
\label{thm:persistent}
    Suppose (DUM) and \eqref{eq:info-persist-pt} hold. 
    Assume the initial information $U_0$ is sufficient for mean outcomes in the following sense:  
    \begin{equation}
    \label{eq:info-persist}
        E[ Y_1(d_1) \mid U_0, Y_0(d_0)] = E [ Y_1 (d_1) \mid U_0] \text{ for all }
        d_0, d_1 \in \{0,1\}.
    \end{equation}
    Then the parallel trends (PT) condition is satisfied. 
\end{example}

Condition \eqref{eq:info-persist} posits that the initial information $U_0$ is sufficient for mean potential outcomes. 
In other words, once conditional on $U_0$, observing the realized past outcome $Y_0 (d_0)$ adds no new information regarding the mean of $Y_1(d_1)$. 
Also, recall the initial information subsumes the sequence of treatment costs $K_t(\cdot)$ (e.g., tuition for a job training program is predetermined and publicly announced as part of $U_0$).  
Therefore, under \eqref{eq:info-persist}, the treatment $D_t$ in each period $t$ is a function of $U_0$ alone. 

The proof of \autoref{thm:persistent} shows that \eqref{eq:info-persist-pt} is in fact stronger than necessary for parallel trends. 
It is also worth mentioning that the sufficient conditions for parallel trends in \autoref{thm:persistent} are consistent with selection on individual fixed effects in $U_0$.

\subsection{Learning and Experimentation}
\label{sec:learning}

We now investigate whether parallel trends can hold if the restrictions on information in \autoref{thm:persistent} are relaxed to allow for dynamic learning and rational forward-looking behavior in treatment choices.
That is, selection into treatment is allowed to depend on information accrued over time, such as past outcomes $Y_0(d_0)$.  
This information is valuable to decision-makers if it is informative about the mean potential outcome under (no) treatment.  

Two effects related to learning are introduced in this subsection. 
First, forward-looking individuals internalize the value of experimentation while choosing treatment in period 0. 
Second, upon observing information from the past outcome in period 0, individuals update beliefs before choosing treatment in period 1. 

It is helpful to conceive of an individual's status in each period (treated/untreated) as ``arms'' (risky/safe) in a multi-armed bandit model.
There is some learning about the mean returns from pulling these arms. 
That is, the decision-makers update beliefs based on past realized outcomes.

We start with a restriction that there is only learning from and about the treated arm. 
In other words, the mean return of the untreated arm is known to decision-makers at the outset, and pulling the untreated arm yields no information about the treated arm. 
This corresponds to a simple two-armed bandit model with a safe and a risky arm. 
The absence of treatment represents the status quo, and the treatment represents a new action (e.g., productivity in a new job, firm entry into a new market, or the effects of a new policy or a new drug).
Note that in the case of a pre-treatment period, this example would assume there is essentially no learning from the past outcome.

\begin{example}[Learning on the Treatment Arm]
    \label{thm:learning}

Suppose (DUM) and \eqref{eq:info-persist-pt} hold. Assume:

\noindent (i) there is no learning across different arms, or from the untreated arm: 
\begin{equation}
\label{eq:control-no-learn}
    E[Y_{1}(d_{1})\mid Y_0(d_0),U_0 ]=E[Y_{1}(d_{1})\mid U_{0}] 
    \quad \text{ when } 
    \begin{cases}
        d_{0} \neq d_{1},  \\ or \\
        d_0 = d_1 = 0;  
    \end{cases}
\end{equation}
\noindent (ii) the treated outcome provides no information about the untreated outcome in period 0, conditional on initial information:
\begin{equation}
\label{eq:no-feedback-safe}
E[Y_0 (0) \mid Y_0(1),U_0] = E [Y_0 (0) \mid U_0].
\end{equation}
Then the parallel trends (PT) condition is satisfied. 
\end{example}

Condition \eqref{eq:control-no-learn} relaxes the sufficiency of initial information in \eqref{eq:info-persist} by allowing:
\[
    E[Y_1 (1) \mid Y_0(1), U_0 ] \neq E[Y_1 (1) \mid U_0].
\]
This introduces net continuation value $W_1 (1) - W_1 (0) \neq 0$ into the decision-maker's problem even when treatment costs $K_t (\cdot)$ are a function of the contemporary choice $d_t$ alone.
Condition \eqref{eq:no-feedback-safe} requires mean independence between the outcomes from two treatment arms within period $t=0$ conditional on initial information $U_0$.
This mean independence is conditional on $U_0$, and does not rule out \textit{unconditional} correlation in the arms across individuals.% 
\footnote{
    For the canonical case with a pre-treatment period (when $P(D_0=0)=1$), \eqref{eq:no-feedback-safe} is not needed for the parallel trends condition.}
For example, in the case of treatment effects of a new drug, the condition requires the patient's period 0 outcome under the new drug to be mean independent of the period 0 outcome under the status quo, conditional on the patient's (private) initial information.

Our next (counter-)example illustrates how learning about the untreated (control) arm leads to violation of parallel trends.
To focus on the selection issue, we consider the case where potential outcomes are binary, and there is a pre-treatment period $P(D_0 =0)=1$.%
\footnote{
    By \autoref{thm:pt-dc}, the same kind of violation of the parallel trends condition would arise conditional on the treatment choice in period 0 if we allowed for a fuzzy design.}

To highlight the impact due to relaxing \eqref{eq:control-no-learn}, we continue to maintain \eqref{eq:info-persist-pt}, i.e., $Y_1 (0)-Y_0(0)$ is mean-independent from $U_0$.
Also, with $D_0 = 0$, the treatment cost at $t=1$ is only indexed by $d_1$; we let $\tilde{K}_1 \equiv K_1 (1) - K_1 (0)$ denote the net cost of treatment in period 1.  

Let $\Omega_{vl}$ denote the set of ``valuable learners'', i.e., the subset of the sample space 
whose optimal decision in period 1 depends counterfactually on the realized outcome in period 0:  
\begin{equation}
    \label{eq:vl}
    \Omega _{vl} \equiv 
    \{ \omega \in \Omega: E[Y_{1}(0)| U_{0} (\omega),Y_{0}(0)=0]\leq
    E[Y_{1}(1) -\tilde{K}_{1}| U_{0} (\omega)]<E[Y_{1}(0)| U_{0} (\omega),Y_{0}(0)=1]\}
\end{equation}
where $U_0 (\omega) $ is parameterized by elements of the sample space. 
Thus, the decision-makers in $\Omega_{vl}$ could learn about the return from the control arm by observing past outcomes. 
In turn, such learning is valuable in period 0 only if it affects decisions in period 1. 

The following example also relaxes the restrictions on learning in \autoref{thm:learning}, i.e., \eqref{eq:control-no-learn}, by allowing learning on the control arm. 
In this case, the parallel trends condition essentially rules out the possibility of valuable learning.  

\begin{example}[Learning on the Control Arm]
    \label{thm:learning-negative}

Suppose (DUM) and \eqref{eq:info-persist-pt} hold. Assume there is a pre-treatment period, i.e., $P(D_0 = 0)=1$, and potential outcomes are binary $Y_t (d_t)\in\{0,1\}$.
Replace \eqref{eq:control-no-learn} with the following conditions:

\noindent (i) there is no correlated learning across no treatment and treatment arms by decision-makers:
\begin{equation}
    \label{eq:no-learning-across}
    E[Y_{1}(1)| Y_0(0),U_0] =E[Y_{1}(1)| U_{0}];
\end{equation}

\noindent (ii) the past outcome of the control arm provides an informative signal about the mean return of that arm in the next period:
\begin{equation}
    \label{eq:cx-learn-signal}
    E [ Y_1 (0) | U_0, Y_0 (0) = 0] \leq E [ Y_1
    (0) | U_0 ] \leq E [ Y_1 (0) | U_0, Y_0 (0) = 1]. 
\end{equation}
Then the parallel trends (PT) condition holds if and only if either:
\begin{enumerate}
    \item there is zero probability of valuable learning, $P ( \Omega_{vl} ) = 0$, or  
    \item valuable learning occurs where untreated outcomes are identical over time almost surely (i.e., $P (Y_0 (0) = Y_1 (0) \mid \Omega_{vl} ) = 1$), and the mean trend of untreated potential outcomes conditional on initial information is zero (i.e. $\tau = 0$ in \eqref{eq:info-persist-pt}). 
\end{enumerate}
\end{example}

If learning about the control arm from past outcomes is valuable to a decision-maker, then past realizations by definition affect future decisions. 
Yet, conditioning on past untreated outcomes introduces backward-looking dynamic selection, similarly to what we saw in \autoref{ex:violate-bs} before.
Thus, the parallel trends condition is violated if there is a positive probability of valuable learning on a subset of the sample space where untreated outcomes are not identical over time almost surely. 

It is noteworthy that a \textit{partial} parallel trends condition as defined in \eqref{eq:pt-dc1} holds in period 0 even when there is learning in period 1 and selection into treatment in period 0, i.e., $P(D_0 \neq 0)>0$.
Namely, in this case, the optimal decision rule in period 0:
\[
    D_0 = \mathbf{1} \{ E [V_0(1) - V_0 (0) + \beta (W_1 (1) - W_1 (0)) \mid U_0] \geq 0 \} 
\]
is still a function of initial information $U_0$.
Therefore, combining with \eqref{eq:info-persist-pt}, similar reasoning as in \autoref{sec:persist-sep} yields:
\begin{align*}
    E [ Y_1 (0) - Y_0 (0) \mid D_0 = d_0]  =  E [ E [ Y_1 (0) - Y_0 (0) \mid U_0] \mid D_0 = d_0 ] = \tau. 
\end{align*}
This implies partial parallel trends \eqref{eq:pt-dc1} from the standpoint of period 0.

Intuitively, decision-makers can anticipate the expected value of future information and even internalize the endogenous motive for learning in their period 0 decisions, which in turn can affect the selection of initial information $U_0$ into treatment in period 0 relative to the case where there is no initial uncertainty about the control arm.
Nevertheless, the important distinction is that the period 0 decisions cannot condition differentially on period 0 versus period 1 potential outcomes, since both outcomes are identically uncertain at the time of decision-making.
In contrast, future decisions can condition on past outcomes, and may optimally do so when past outcomes convey valuable information about the future (which would generally invalidate the parallel trends condition).
In summary, learning about the control arm in our example violates parallel trends due to selection on past untreated outcomes, rather than the forward-looking value of strategic experimentation.    

\subsection{Selection on Present Outcomes}
\label{sec:present}

Next, we explore the case when treatment is a function of present outcomes. 
To begin, consider a Roy model with treatment-invariant uncertainty, which is subsumed in our (DUM) context, with $K_t (\mathbf{d}^t) = 0$, and $ U_1 (0) = U_1 (1) \equiv U_1$. 
In other words, such a model allows incremental information $U_1(d_0)$, but either rules out experimentation (i.e., $Y_0(d_0)$ is not subsumed in $U_1(d_0)$, since $U_1(d_0)$ does not vary with $d_0$), or requires $Y_0(d_0)$ to be independent from, and thus not indexed by, potential treatment in period 0.

Throughout \autoref{sec:present}, we no longer maintain the mean independence of untreated trends $Y_1(0) - Y_0(0)$ from the initial information $U_0$ as in  \eqref{eq:info-persist-pt}.
This is because our focus in this subsection is on the endogenous selection of treatment due to \textit{present} potential outcomes, rather than on the evolution of information.
 
\begin{example}[Static, Repeated Roy Model]   \label{ex:roy}

Recall the model in (DUM) with $\mathcal I_0 = \{U_0\} $ and $\mathcal I_1(d_0) = \{U_0, U_1(d_0) \}$. Suppose $Y_t(d_t) \in \{0,1\}$ for $d_t=0,1$, and $P(Y_t(1) \geq Y_t(0)) \in (0,1)$ for $t=0,1$. Assume: \medskip

(i) the initial information $U_0$ subsumes $Y_0(\cdot)$,  and $ K_t(\cdot)=0$ for $t=0,1$; \medskip

(ii) the incremental information accrued before period 1 consists of both potential outcomes in period 1, i.e. $U_1(d_0) = U_1 \equiv \{Y_1(0),Y_1(1)\}$ for $d_0=0,1$. \medskip

\noindent Under these conditions, there is no net continuation value, because $W_1 (1) = W_1 (0)$. 
Therefore, the optimal treatment rule is reduced to $D_t = \mathbf{1} \{ Y_t (1) \geq Y_t (0) \}$. \medskip

The parallel trends (PT) condition holds if and only if the following conditions hold jointly: 
    \begin{enumerate}
        \item \label{item:stat-roy}
        Untreated outcomes are stationary: $E [ Y_0 (0) ] = E [ Y_1 (0) ];$
        \item \label{item:as-roy}
        Untreated outcomes are degenerate at 1 among the ever-untreated:
        \[
            P(Y_0 (0) = Y_1 (0) = 1 | D_0 D_1 = 0 ) = 1.
        \]
    \end{enumerate}
\end{example}

The claim in \autoref{ex:roy} follows from the simple treatment rule. 
This example shows that the parallel trends condition restricts the distribution of potential outcomes when the treatment is determined by a comparison of present potential outcomes.
For example, to be consistent with parallel trends in this example, the probability of strict switches in potential outcomes across time must be zero:
    \[
            P (Y_t (0) > Y_t (1), Y_{1-t} (0) < Y_{1-t} (1)) = 0 \text{ for } t=0,1.
    \]
Such requirements rule out non-trivial independence of potential outcomes over time. 

Next, we generalize the treatment rule to an arbitrary function of present information. 
We focus on the case where selection in each period is based on serially uncorrelated factors. 
In this case, the parallel trends condition is reduced to strict exogeneity of untreated outcomes.

\begin{proposition}
\label{thm:ind-exogeneity}
    Suppose in each period, decision-makers choose treatment $D_t$ as a deterministic function $f_t(\cdot)$ of history-invariant information $U_t$ about present outcomes $Y_t (\cdot)$:
    \begin{equation}
    \label{eq:selection-present}
        D_t = f_t(U_t). 
    \end{equation}
    If potential outcomes and information are independent over time: 
    \begin{equation}
        \label{eq:uy-ind}
        (U_0, Y_0 (\cdot)) \perp (U_1, Y_1 (\cdot)),
    \end{equation}
    then a necessary and sufficient condition for parallel trends (PT)  is strict exogeneity of untreated outcomes:
    \begin{equation}
        \label{eq:strict-exogeneity}
        E [ Y_t (0) \mid D_0 = d_0, D_1 = d_1] = \mu_t 
    \end{equation}
    for some constants $\mu_t \in \mathbb{R}$ and for all $t$, all $(d_0,d_1)$ occurring with positive probability.
\end{proposition}

The combination of \eqref{eq:selection-present} and \eqref{eq:uy-ind} is testable by comparing observed untreated outcomes across treatment sequences. For example: 
\[
    E [ Y_0 \mid D_0 = 0, D_1 = 0] = E [ Y_0 \mid D_0=0, D_1 =1].
\]
The role of parallel trends is then to extrapolate the equality to mean untreated potential outcomes $Y_1(0)$ in period 1.

\subsection{Staggered Designs}
\label{sec:optimal-stopping}

A common DiD setting involves \textit{staggered designs}, where treatments are irreversible, i.e., $D_1 \geq D_0$.
We investigate the viability of parallel trends (PT) under two micro-founded models of staggered designs. 
The first revisits the repeated Roy setup in \autoref{ex:roy}, but with prohibitive costs for treatment reversal $K_1(1,0) = \infty$. 
This introduces dynamic motives for treatment selection (with \textit{nonzero} net continuation value) so that the staggered design arises endogenously within the DUM context. 
The second model rationalizes the sequence of irreversible treatments as a solution to an optimal stopping problem.

We relate these models to the repeated Roy model in \autoref{ex:roy} and the models with learning and experimentation in \autoref{thm:learning} and \autoref{thm:learning-negative}.
Our goal is to understand how these staggered designs alter the implications of dynamic selection for parallel trends.

\subsubsection{DUM with Prohibitive Costs for Treatment Reversal}

In the repeated Roy model in \autoref{sec:present}, we abstracted away from history-dependent treatment costs $K_0(d_0)$ and $K_1(d_0,d_1)$ by assuming that these are finite and subsumed in the initial information $U_0$, and by normalizing them to zero. 
This partially led to the lack of history-dependent dynamics of information in \autoref{ex:roy}.

In contrast, the next example rationalizes irreversible treatments using prohibitive costs for a reversal in treatment (i.e. when $K_1(1,0)$ is large), and examines its implications on the parallel trends condition.

\begin{example}[Roy Model with Prohibitive Costs for Treatment Reversal]
\label{thm:strong-roy-os} 
    Recall the model in (DUM). Suppose that $\mathcal I_0 = \mathcal I_1(d_0) = \{U_0\}$, with $U_0$ subsuming $Y_t(\cdot)$ and $K_t (\cdot)$ for $t=0,1$, and that $Y_t(d_t) \in \{0,1\}$ for $d_t=0,1$.
    Let $K_1 (1,0) = \infty$ (infinite cost for treatment reversal) while $K_t(\mathbf{d}^t) = 0$ otherwise.
    
    Under these conditions, the optimal decision rule at $t=0$ is:%
    \footnote{
        At $t=0$, the payoff from $ D_0 = 0 $ is $ Y_0(0) + \beta \max\{Y_1(1), Y_1(0)\} $ while that from $D_0=1$ is $ Y_0(1) + \beta Y_1(1)$.
        (The latter expression is simplified because $K(1,0)=\infty$ implies: if $D_0=1$, then the optimal continuation value can not involve $D_1=0$.)  
        Taking the difference of these two expressions leads to the optimal treatment rule in \eqref{eq:roy-os1}. }
 \begin{align}
     \label{eq:roy-os1}
     D_0 = \mathbf{1} \{ Y_0 (1) - Y_0 (0) + \beta \min \{ Y_1 (1) - Y_1 (0), 0  \} \geq 0\}.  
 \end{align} 
Suppose $P(D_0 = D_1 = 0), P (D_0 < D_1) \in (0,1)$.  
Then the parallel trends condition (PT) holds if and only if Conditions 1 and 2 of \autoref{ex:roy} hold jointly.%
\footnote{
In Condition 2, note that the event of being ever-untreated can differ across Examples \ref{ex:roy} and \ref{thm:strong-roy-os}.
}
\end{example} 

Unlike in the Roy model of \autoref{ex:roy}, the prohibitive cost for treatment reversal in \autoref{thm:strong-roy-os} introduces dynamic considerations through the nonzero net continuation value:
 \begin{equation}
     \label{eq:continuation-os}
     W_1 (1) - W_1 (0) = \min \{ Y_1 (1) - Y_1 (0) , 0 \} \leq 0,
 \end{equation}
which measures the foregone option value of choosing $D_1=0$. 
Thus, in contrast to the examples with learning, expectations about future outcomes may influence (and, in the present case, discourage) treatment in period 0 even without information acquisition considerations. 

To relate the logic behind \autoref{thm:strong-roy-os} to that of \autoref{ex:roy}, it is useful to decompose the population in terms of their (now counterfactual) decisions without the prohibitive costs for treatment reversal.   
Those who would have been movers out of treatment $(D_0 > D_1)$ when the treatment reversal cost is zero now become either always- or never-treated.
Those who become never-treated as a result of the treatment reversal costs are those with a relatively high net value of no treatment in period 1.  
Those who would have been never-treated even without the treatment reversal costs remain such and have a zero untreated trend by the same logic of \autoref{ex:roy}. 
Since movers into treatment 
have a high value of no treatment in period 0, parallel trends between the (constrained) never-treated and movers into treatment is satisfied if and only if the untreated trend is zero and untreated outcomes among the ever-untreated are degenerate at 1.  

\subsubsection{Optimal Stopping}
\label{sec:os}
Next consider a simple, two-period optimal stopping problem. Let $D_t = 1$ denote an irreversible decision to ``stop'', which leads to a terminal utility that is normalized to zero. 
For example, a firm may choose to irreversibly exit a market and only collect the scrap value of its assets.%
\footnote{
    Optimal stopping has a long history in empirical economics. Early examples are the two classic works by \cite{mortensen1970job} and \cite{lippman}.
    Other important contributions in economics are the works of \cite{pakes1986patents} , \cite{rust1987optimal}, and \cite{wolpin1987}.
    More generally, this problem is part of the dynamic Markovian decision problem literature that includes reinforcement learning
    problems.
}
In each period $t$, the decision to ``continue''  ($D_t = 0$) incurs a continuation cost $K_t$, and leads to an instantaneous payoff $Y_t(0)$. 
Let the initial information available to the decision-maker be $U_{0}\equiv (K_{0},K_{1},\zeta )$, with $\zeta $ denoting any feature that is potentially correlated with the transition of potential outcomes $Y_t(0)$ under continuation.

At the start of period $t=1$, a decision-maker has accrued information $\mathcal I_1(0) = \{U_0,U_1(0)\}$ with $U_1(0) = Y_0(0)$, and chooses:%
\begin{equation}
D_{1}(0)=\mathbf{1}\{E[Y_{1}(0)\mid Y_{0}(0),U_{0},D_{0}=0]-K_{1}\leq 0\}\text{,}  \label{D1}
\end{equation}%
with a value function 
\[
\nu_{1}(Y_0(0))\equiv \max \{0,E[Y_{1}(0) \mid Y_{0}(0),U_{0},D_{0}=0]-K_{1}\}\text{.}
\]%
Note we suppress $U_0$ as an argument in $\nu_1(\cdot)$ for simplicity. 
At the start of period $t=0$, a decision-maker has initial information $\mathcal I_0 \equiv \{U_0\}$ and chooses: 
\begin{equation}
D_{0}=\mathbf{1}\{ E [ Y_{0}(0) + \beta \nu_{1}(Y_0(0)) \mid U_{0}] -K_{0}\leq 0 \} \text{.}
\label{D0}
\end{equation}%
Thus $D_{0}$ is determined by $U_{0}$ while $D_1$ is determined by $Y_0(0)$ and $U_0$.

For the rest of \autoref{sec:optimal-stopping}, we maintain the mean independence condition in \eqref{eq:info-persist-pt}, 
\begin{equation*}
      E[Y_{1}(0)-Y_{0}(0)\mid U_{0}]=\tau \text{.}
\end{equation*}%
We show that, even with such mean independence in the trend of $Y_t(0)$, the parallel trends condition fails generically because of learning and forward-looking motives in the selection into treatments. 

Similar to what we have done for the general dynamic utility maximization (DUM) model in \autoref{sec:persist-sep} and \autoref{sec:learning}, we now analyze the conditions for parallel trends in the optimal stopping problem with or without substantive learning from past outcomes.

\bigskip

\noindent \textbf{Case 1 (Sufficient Initial Information).} 
Suppose \eqref{eq:info-persist-pt} holds. Assume 
\begin{equation}
E[Y_{1}(0)\mid U_{0},Y_{0}(0)]=E[Y_{1}(0)\mid U_{0}]\text{,}
\label{assum:one}
\end{equation}%
so that $D_{1}$ is a function of $U_{0}$ alone (just as $D_0$ is). 
By the Law of Iterated Expectation,
\[
E[Y_{1}(0)-Y_{0}(0)\mid D_{0}=1] = 
E[Y_{1}(0)-Y_{0}(0)\mid D_{0}=0,D_{1}=d]=\tau \text{ for d = 0,1.} 
\]%
Therefore, the parallel trends (PT) condition holds.

\bigskip

\noindent \textbf{Case 2 (Learning from Past Outcomes).}
Suppose (\ref{eq:info-persist-pt}) holds. Relax (\ref{assum:one}) so that 
$E[Y_{1}(0)\mid U_{0},Y_{0}(0)]$ is non-degenerate in $Y_{0}(0)$.
The choice/treatment rule at $t=0$ remains the same as in \eqref{D0}, which indicates $D_0$ only depends on $U_0$. Thus, we can write the treatment rule at $t=1$ in \eqref{D1} as:%
\[
D_{1}(0)=\mathbf{1}\{E[Y_{1}(0)\mid Y_{0}(0),U_{0}]-K_{1}\leq 0\}%
\text{.} 
\]%

For the rest of this subsection, we present and discuss a necessary and sufficient condition for parallel trends in Case 2.
Let $\mathcal{U}^d$ denote the set of realized values of the initial information $U_0$ that lead through \eqref{D0} to $D_{0}=d$. 
For any $u \in \mathcal U^0$, let $\mathcal{S}^{d}(u)$ denote the set of realized values $y$ for $Y_0(0)$ such that $(u,y)$ induces the event ``$D_{0}=0$ and $D_{1}(0)=d$''. 

With treatments determined by optimal stopping, the parallel trends (PT)  condition is: 
\begin{equation*}
E[Y_1(0)-Y_0(0)\mid D_0=1] = E[Y_1(0)-Y_0(0)\mid D_0=0,D_1=d] \text{ for } d = 0,1.    
\end{equation*}
By the mean independence of $Y_1(0)-Y_0(0)$ from $U_0$ in (\ref{eq:info-persist-pt}),
\[
E[Y_{1}(0)-Y_{0}(0)\mid D_{0}=1]= (P_{1,\cdot})^{-1} \times \int_{\mathcal{U}^1} E[Y_{1}(0)-Y_{0}(0)\mid U_{0}=u] dF_{U_{0}}(u)=\tau, 
\]
where $P_{1,\cdot} \equiv P(D_0=1) = P(U_0\in \mathcal U^1)$. 
Furthermore, for $d=0,1$,
\begin{eqnarray}
\label{eq:pt-part-two}
E[Y_{1}(0)-Y_{0}(0)\mid D_{0} = 0,D_{1}=d] = (P_{0,d})^{-1} \times \int_{\mathcal U^0} \delta_d(u)dF_{U_0}(u), \end{eqnarray}%
where 
\[ 
\delta_d(u)\equiv \int_{\mathcal{S}^d(u)} \{E[Y_{1}(0) \mid U_{0}=u,Y_{0}(0)=y]-y\}dF_{Y_0(0)|U_0=u}(y),
\] 
and $ P_{0,d} \equiv P(D_0=0,D_1=d) = P(U_0\in \mathcal U^0,Y_0(0) \in \mathcal S^d(U_0))$.
As a result, in Case 2, the parallel trends condition is equivalent to the statement that the right-hand sides of \eqref{eq:pt-part-two} equal $\tau$ for $d=0,1$. 
Furthermore, under \eqref{eq:info-persist-pt}, 
$ \delta_1(u) + \delta_0(u) = E[Y_1(0) - Y_0(0)\mid U_0=u] = \tau$.
It then follows that parallel trends hold if and only if 
\begin{equation} 
    \label{eq:pt-ns-opt-stop}
    \int_{\mathcal U^0} \delta_0(u) dF_{U_0}(u) = \tau \times \mathcal P_{0,0}.    
\end{equation}

This necessary and sufficient condition for parallel trends in Case 2 fails generically if there are no further restrictions on the joint distribution of $(Y_1(0),Y_0(0),U_0)$ beyond \eqref{eq:info-persist-pt}. 
To see this, consider a simple example $Y_1(0) = Y_0(0) + \eta$ with 
\[   E[\eta\mid Y_0(0) \in \mathcal S^d(u), U_0=u] \equiv \varphi_d(u)  \] 
for all $u\in\mathcal U^0$.
By construction, $ \delta_d(u) = \varphi_d(u) P_d(u)$ for $u\in \mathcal U^0$, where $P_d(u) \equiv P(Y_0(0)\in \mathcal S^d(u) \mid U_0=u) $.
In this model, mean independence in \eqref{eq:info-persist-pt} imposes
\[
    E[\eta\mid U_0=u] = \varphi_1(u)P_1(u) + \varphi_0(u)P_0(u) = \tau \text{ for all } u,   
\]
but does not further restrict how $\varphi_d(\cdot) $ varies between $d=0,1$ over $\mathcal U^0$. 
As a result, the necessary and sufficient condition for parallel trends in \eqref{eq:pt-ns-opt-stop} is not guaranteed to hold, because the left and right sides of \eqref{eq:pt-ns-opt-stop} differ generically without further restrictions:
\[
    \int_{u\in\mathcal U^0} \delta_0(u) dF_{U_0}(u) = 
    \int_{u\in\mathcal U^0} \varphi_0(u)P_0(u) dF_{U_0}(u) \neq \int_{u\in\mathcal U^0} \tau P_0(u) dF_{U_0}(u) = \tau \times \mathcal P_{0,0}.
\]

\noindent This leads to failure of (PT) to hold in optimal stopping problems when we allow for learning from past outcomes  (even while maintaining the mean independence restriction \eqref{eq:info-persist-pt}).
 
\subsection{Discussion: Past Treatment, Anticipation, and Latent Factors} 
\label{sec:extra}

So far, our overarching dynamic utility maximization framework (DUM) and specific examples have focused on selection into treatment given potential outcomes indexed by treatment in the same period.  
We now briefly discuss extensions of the outcome model (\autoref{assn:present}) and their relation to our preceding analysis.
We maintain our focus on individual decision problems, thereby abstracting from spillovers across units such as social learning or general equilibrium effects.  

One extension of our model relaxes \autoref{assn:present} to allow potential outcomes to vary by treatment \emph{sequence}, i.e., $Y_t (d_0, d_1)$, thereby introducing the possibility of dynamic effects. 
In that case, the parallel trends assumption is typically expressed in terms of the trend of never-treated potential outcomes:
    \begin{equation}
        \label{eq:pt-g}
        E [ Y_1 (0,0) - Y_0 (0,0) | D_0 = d_0, D_1 = d_1] = \tau
    \end{equation}
for some constant $\tau \in \mathbb{R}$ and all $d_0, d_1 \in \{0,1\}$.% 
\footnote{
    In the case of staggered designs $D_1 \geq D_0$, the treatment sequence (and
    thus the generalized potential outcome parameterization) is summarized by
    the timing of first treatment $G$ with realizations $g \in \{0,1,\infty\}$, where
    $g = \infty$ denotes the never-treated sequence $d_0 = d_1 = 0$ in the basic
    two-period data. 
    In that case, condition \eqref{eq:pt-g} coincides with a two-period
    version of Assumption 1 of \cite{sundid}. 
}

In our DUM model, the dependence of future outcomes $Y_1 (d_0, d_1)$ on past treatment $d_0$ introduces an additional motive for forward-looking behavior and selection into treatment in period 0 based in part on persistent effects in period 1, as in the preceding case of optimal stopping.
    
The dependence of past outcomes $Y_0 (d_0, d_1)$ on future treatment $d_1$ also allows anticipation of (known) treatments to affect potential outcomes.%
\footnote{
    It is useful to distinguish this from anticipation in the treatment decision based on expectations about future outcomes and treatments, which is compatible with our baseline specification. 
    Such anticipation in treatments could also induce a kind of anticipation in \emph{realized} outcomes $Y_0$, which we showed in \autoref{sec:learning} need not violate parallel trends per se. }
However, typical DiD methods require significant restrictions on anticipation in order to identify an ATT.
For example, identification of such an effect among switchers into treatment $D_0 < D_1$ requires assuming no anticipation on average among the switchers: 
    \begin{equation}
        \label{eq:no-anticipation}
        E [ Y_0 (0,0) | D_0 < D_1 ] = E [ Y_0 (0,1) | D_0 < D_1] 
    \end{equation}
in order to identify the never-treated outcome in period 0 among switchers in \eqref{eq:pt-g}.
Stated as such, no anticipation is partially an assumption about selection based on future treatment. 
In our DUM model where $Y_0$ is realized before $D_1$ is chosen, no anticipation is most sensible in the stronger guise where $Y_0 (0,d_1)$ is everywhere constant in $d_1$. 
Alternatively, anticipation could occur based on uncertain future treatments\footnote{This anticipation effect is certainly present also in macro where current outcomes (and decisions) by the Fed anticipate and are affected by future potential decisions.}, as in \cite{malani2015anticipation}.
For example, in our DUM model, uncertainty about future treatment arose with learning about expected outcomes from period 0.  
We embed this case in a more general discussion about our implicit assumptions regarding the determination of potential outcomes below. 
    
More generally, potential outcomes may be affected by other latent, unobserved choices, which in turn may depend on the expected sequence of treatment. 
For concreteness, consider again the random coefficient model \eqref{eq:outcome-model} with direct dependence on the present treatment, but suppose that treated and untreated outcomes are also affected by an additional latent effort $f$ chosen by decision-makers before treatment in each period. That is,
    \begin{equation}
        \label{eq:effort}
        Y_{it} (d_{it}) = \alpha_i + \delta_t + \beta_{it} d_{it} +
        \gamma_{it} f_{it} + \rho_{it} d_{it} f_{it} + \varepsilon_{it}
    \end{equation}
where $f_{it}$ is an endogenous level of effort also chosen by decision-makers.% 
\footnote{
    This extension is similar in spirit to a classic literature on the estimation of production functions with endogenous inputs (\cite{marschakandrews1944}) using panel data (\cite{hoch62, mundlak63}), except that \eqref{eq:effort} allows for an interaction term between effort and treatment and we think of effort as being unobserved. 
    If the additional choice variable is instead observed, the preceding literature suggests more structural proxy-based approaches to control for unit-level heterogeneity, e.g., \cite{olleypakes96, levinsohnpetrin03}.}
(In keeping with notation above, we suppress dependence of $Y_{it}(\cdot)$ on this endogenous input effort.)
Taking time trends $\delta_t$ to be non-stochastic, and assuming the effort levels $F_{it}$ are \textit{unobserved} by the researcher, one can write the parallel trends condition as:
\[
    E [ \gamma_{i1} F_{i1} - \gamma_{i0} F_{i0} + \varepsilon_{i1} -
    \varepsilon_{i0} | D_{i0} = d_0, D_{i1} = d_1] 
\]
is constant across $d_0, d_1 \in \{ 0,1 \}$. 
In addition to supposing that the difference in errors $\varepsilon_{it}$ is mean independent of treatment \eqref{eq:demit}, a sufficient condition for parallel trends is that there is no selection on the difference in untreated returns from effort: 
\begin{equation}
    \label{eq:diff-returns-effort}
    E [ \gamma_{i1} F_{i1} - \gamma_{i0} F_{i0} | D_{i0} = d_0, D_{i1} = d_1]
    = \psi 
\end{equation}
for some constant $\psi \in \mathbb{R}$ and all $d_0, d_1 \in \{0,1\}$.
For example, this would hold if the marginal return from effort $\gamma_{it}$ is constant across units and time, and the change in effort $F_{i1} - F_{i0}$ is independent of treatments $(D_{i0}, D_{i1})$. 
However, the latter seems especially unlikely when effort also affects the returns from treatment, $\rho_{it} \neq 0$. 
There is also a further issue of causal interpretation, since even the difference in potential outcomes holding effort fixed at $f_{it}=f$ contains the returns from effort, $ Y_{it} (1) - Y_{it} (0) = \beta_{it} + \rho_{it} f $.
If the interest is in identifying (averages of) $\beta_{it}$, then  this setting requires methods beyond the TWFE paradigm that is the focus of this paper, and so we do not consider the problem further. 
For the same reason, we do not consider lagged outcomes studied by a large literature on dynamic panel data models [e.g., \cite{andersonhsiao82,bhargavasargan83, arellanobond91}]. 
Instead, we return to the canonical setup (\autoref{assn:present}) and consider alternative methods for identification when multi-period parallel trends may be violated.

\section{Alternative Identification without Parallel Trends}
\label{sec:id} 

Based on insights from the models in \autoref{sec:examples}, this section considers alternative approaches for identification when the parallel trends condition fails. 
In developing these alternative approaches, we emphasize that there is typically not a single solution because different models of the data-generating process yield different results. 
This is similar to the framework of \cite{manski2018rtc}.%
\footnote{See also \cite{rothrambachan} for recent work in this direction.}  

We focus on a single target parameter, the average treatment effect at $t=1$ among switchers into treatment: 
\begin{equation}
    \label{eq:te-switchers}
    E [ Y_1 (1) - Y_1 (0) \mid D_0 < D_1]
\end{equation}
The treatment effect among switchers \eqref{eq:te-switchers} coincides with the usual treatment effect on the treated in period 1, i.e., $E [Y_1 (1) - Y_1 (0) | D_1 = 1]$, in the case with a pre-treatment period where $P(D_0 = 0)=1$.
The parameter defined in \eqref{eq:te-switchers} provides a natural generalization under fuzzy designs, where $P(D_0 = 1) > 0$.%
\footnote{
    Alternatively, $E[Y_1 (1) - Y_1 (0) \mid D_1 = 1]$ in the fuzzy design would average over treated and untreated observations in period 0, whereas $E [ Y_1 (1) - Y_1 (0) \mid D_0 = D_1 = 1]$ would likely require stronger assumptions, since among this subgroup the untreated outcome is never observed.
}
For example, the treatment effect among switchers also serves as a building block for the DID-M estimator proposed in \cite{CD}.

The treatment effect parameter defined in \eqref{eq:te-switchers} differs qualitatively from the LATE with time as an instrument, despite a seeming resemblance.
In typical panel data settings, time is not randomly assigned or excluded from the potential outcome equation, and treatment may not be monotonic in time.%
\footnote{
    For an analysis where time is a monotonic instrument, see \cite{de2018fuzzy}.
}
Additionally, treatment is observed in each time period, whereas it is typically only observed for one realization of instrument values.
We return to the conceptual relations between the two parameters and their identification in \autoref{sec:stat}.
Throughout we restrict ourselves to settings where a positive mass of never-treated and switchers into treatment exists, $P(D_0=D_1=0)>0, P(D_1 > D_0) > 0$.

\subsection{Partial Parallel Trends} 

One takeaway from the models in \autoref{sec:examples} is that parallel trend comparisons across some treatment sequences may be more robust to selection concerns than others. 
This motivates the question: what can be identified under relaxations of the full parallel trends (PT) condition that impose parallel trends only among subsets of the treatment sequence realizations?

First, recall the following partial parallel trends assumption from \autoref{thm:pt-dc}, which relaxes full parallel trends by only imposing parallel-trend restrictions between switchers into treatment and the never-treated.
\begin{assumption}[Partial Parallel Trend Between Switchers and Never-Treated] {\,}
\label{assn:ppt1}
\[
E [ Y_1 (0) - Y_0 (0) | D_0 = 0, D_1 = d_1] = \tau_{0,1} \quad \text{for $d_1 \in \{0,1\}$}
\]
\end{assumption}

\noindent
\autoref{assn:ppt1} only requires that the untreated trend is independent of the period 1 treatment conditional on $D_0  = 0$; it imposes no restriction on trend-based selection in the initial period 0. 
(Of course, this assumption is the same as assuming full parallel trends in a set-up with a pre-treatment period, where there is by definition no selection in period 0.)
\autoref{assn:ppt1} is sufficient for recovering the period 1 average treatment effect for switchers into treatment.

\begin{proposition}
\label{thm:ppt}
Let \autoref{assn:ppt1} hold. Then:
\begin{align}
    &E[ Y_1 (1) - Y_1 (0) | D_0 < D_1] \notag \\
    &\quad= 
    E [ Y_1  - Y_0 | D_0 < D_1] - E [Y_1  - Y_0  | D_0 = D_1 = 0]. 
    \label{eq:te-switchers-into} 
\end{align}    
\end{proposition}

\noindent
The result follows immediately from the usual logic with a pre-treatment period applied to the subgroup where $D_0 = 0$.%
\footnote{
Relating to the existing literature, \autoref{thm:ppt} captures the basic intuition behind the $\text{DID}_{+}$ estimator (in turn underlying the $\text{DID}_\text{M}$ estimator) of \cite{CD} but shows that it holds under a logically weaker assumption than the full parallel trends (PT) condition. This weaker assumption suffices for comparing the evolution of the untreated trend among switchers into treatment (where this trend is counterfactual) and the never-treated (where this trend is observed). 
A similar intuition also underlies the ``building block'' estimators of \cite{callawaysatanna} and \cite{sundid} in staggered settings.
In that case, however, our simple two-period framework does not distinguish between whether the control group $\{D_0 = D_1 = 0\}$ is never-treated or not-yet treated.
For elaboration of the distinction, see \cite{callawaysatanna}.
Of course, it is worth noting that in each of these papers, the primary contribution is to provide methods for aggregating over such basic effects.
} 
Furthermore, it is straightforward to show that \eqref{eq:te-switchers-into} is one of only two basic treatment effects among the family $E[ Y_r (1) - Y_r (0) | (D_0,D_1) = (d_0, d_1)]$ that are identified even under full parallel trends.%
\footnote{
By symmetry, the other is the period 0 average treatment effect among switchers out of treatment, 
\[
E[ Y_0 (1) - Y_0 (0) | D_0 > D_1]
\]}
In summary, violations of the full parallel trends (PT) condition need not lead to failures of point identification of the treatment effect parameter in \eqref{eq:te-switchers} per se.

Further motivated by the more general learning example in \autoref{sec:learning}, we also consider the content of a weakened parallel trend restriction that only conditions on past realized treatments. 

\begin{assumption}[Forward Parallel Untreated Trend] 
\label{assn:fpbt}
\begin{equation}
\label{eq:fpbt}
    E [Y_1(0) - Y_0(0) | D_0 = d_0 ] = \tau_0 
    \quad \text{for $d_0 \in \{0,1\}$.}
\end{equation}
\end{assumption}

\noindent
As shown in \autoref{sec:learning}, such an assumption is consistent with choice environments where decision-makers can learn but have no differential information about untreated outcomes across period 0 and 1 at the time of deciding in period 0.
\autoref{assn:fpbt} relaxes multi-period parallel trends (PT) by only imposing parallel trends conditional on treatment in period 0.
This allows for selection on \emph{past} outcomes in the treatment decision of period 1.%
\footnote{
A common approach in such settings is to use estimators that condition or match on lagged dependent variables (e.g.~\cite{abadie2010synthetic} and \cite{dehejia1999causal}, respectively);
see also \cite{angrist2008mostly} and \cite{ding2019bracketing} for circumstances where the treatment effect is bounded between the DiD and lagged dependent variable estimators.
However, conditions for consistent estimation with lagged dependent variables may be necessarily strong (e.g.~\cite{nickell1981biases}). 
}

Unlike the full parallel trends condition in (PT) and the partial parallel trends condition in \autoref{assn:ppt1}, \autoref{assn:fpbt} does not point identify the average treatment effect among switchers into treatment (and thereby, as a weakening of the parallel trends condition, any other basic treatment effects).  
Consider the identity:
\begin{equation}
    \label{eq:switch-identity}
    E [ Y_1 (1) - Y_1 (0) | D_1 > D_0] = E [ Y_1 (1) - Y_0 (0) | D_1 > D_0] - E [ Y_1 (0) - Y_0 (0) | D_1 > D_0]
\end{equation}
The first term on the right-hand side is identified, and so it remains to identify the second term, namely the untreated trend among the switchers.
Decomposing \eqref{eq:fpbt} into its constituent treatment sequences, \autoref{assn:fpbt} is equivalent to: 
\begin{align*}
    &\sum_{d_1} P (D_1 = d_1 | D_0 = 0) E [ Y_1 (0) - Y_0 (0) | D_0 = 0, D_1 = d_1] \\
    =
    &\sum_{d_1'} P (D_1 = d_1' | D_0 = 1) E [ Y_1 (0) - Y_0 (0) | D_0 = 1, D_1 = d_1'].
\end{align*}
Thus, \autoref{assn:fpbt} only allows us to express the untreated trend in \eqref{eq:switch-identity} in terms of other unobserved untreated trends among the always-treated and switchers out of treatment:
\begin{align*}
E [ Y_1 (0) - Y_0 (0) |D_1 > D_0] 
&=
\frac{P(D_1=1|D_0=1)}{P(D_1=1|D_0=0)} E [ Y_1 (0) - Y_0 (0) | D_0 = D_1 = 1] \\[0.05in]
&+
\frac{P(D_1=0|D_0=1)}{P(D_1=1|D_0=0)} E [ Y_1 (0) - Y_0 (0) | D_0 > D_1] \\[0.05in]
&- 
\frac{P(D_1=0|D_0=0)}{P(D_1=1|D_0=0)} E [ Y_1 (0) - Y_0 (0) | D_0 = D_1 = 0].
\end{align*}
A stark research design that circumvents this issue is one where $P(D_1 = 0) = 1$, but in that case there is no selection in period 1 and so full parallel trends (PT) is trivially recovered.
Next, we consider other assumptions --- in some cases nested by the above --- that allow at least partial identification of the desired target parameter.

\subsection{Mean Stationarity}
\label{sec:stat}

We now revisit the role of mean stationarity of untreated outcomes. 
In \autoref{sec:examples}, mean stationarity arose as a necessary condition for parallel trends in Examples \ref{ex:violate-bs}, \ref{ex:roy}, and \ref{thm:strong-roy-os} and as a salient case of the dynamic choice models with (no) learning in Examples \ref{thm:persistent}, \ref{thm:learning}, and \ref{thm:learning-negative}, where agents learned about the payoffs of fixed bandit arms.
Next we show how, when applicable, such mean stationarity can also bypass selection concerns by providing an alternative path to identification.
Recall the notion of mean stationarity of untreated outcomes: 

\begin{assumption}[Mean Stationarity of Untreated Outcomes] 
    \label{assn:stat}
%     The mean untreated outcome is constant across time:  
    \[
        E [ Y_1 (0)-  Y_0 (0) ] = 0. 
    \]
\end{assumption}

\noindent
\autoref{assn:stat} is not nested by the parallel trends (PT) condition. 
Parallel trends does not restrict the trend to zero, and so it does not imply stationarity.
Conversely, stationarity makes no assumptions about endogeneity of the realized treatment sequence $(D_0,D_1)$, and so it does not imply any version of parallel trends.
In terms of the linear panel data model for potential outcomes in \eqref{eq:outcome-model}, mean stationarity assumes the untreated (and non-stochastic) time trend is constant, $\delta_0 = \delta_1$ (under the location normalization that time-varying errors $\varepsilon_{it}$ have zero means), but it makes no assumption about the exogeneity of treatments.%
\footnote{
    However, when outcomes are affected by an additional unobserved effort $F_{it}$ as in \autoref{sec:extra}, mean stationarity requires constant realized returns from effort
    $E [ \gamma_{i1} F_{i1} - \gamma_{i0} F_{i0} ] = 0$, whereas under parallel trends these returns are allowed to vary over time \eqref{eq:diff-returns-effort}, albeit independently of the treatment sequence. 
}
Furthermore, this does not typically impose restrictions on the trend in realized outcomes because of selection.

Even though the unconditional stationarity in \autoref{assn:stat} does not imply parallel trends, it also identifies the average treatment effect on the treated in the basic setting with a pre-treatment period.% 
\footnote{
A related version of time invariance is considered by \cite{manski2018rtc}. 
The main difference between our assumptions is that theirs is imposed by indexed observation rather than in expectation; in our setting, an elementwise version of time invariance, i.e., $Y_0 (0) = Y_1 (0)$, would also imply parallel trends.
}

\begin{proposition}
    \label{thm:id-tot}
    Suppose $D_0 = 0$. Under Assumption \ref{assn:stat}, 
    \[
        E [ Y_{1} (1) - Y_{1} (0) | D_1  = 1]
        =  E [ Y_{1} (1) - Y_{1} (0) | D_0 < D_1 ]
        =  \frac{E [ Y_1 ] - E [ Y_0 ]}{P (D_1 = 1)}.
    \]
\end{proposition}

The proof is provided in \autoref{apx:proofs}.
Since $D_1 \geq D_0 = 0$, this appears related to the LATE theorem of \cite{imbensandangrist} with time as an instrument.%
\footnote{
For previous results interpreting time as an instrument, see also \cite{de2018fuzzy}.
}
However, recall a few important differences from the previous discussion of the target parameter \eqref{eq:te-switchers}. 
First, the time index violates the exclusion restriction since the potential outcome $Y_t (d)$ can depend on $t$.
Time invariance (\autoref{assn:stat}) can be interpreted as a mean exclusion restriction on the untreated outcomes, but no such assumption is imposed on mean treated outcomes. 
Instead, mean treated outcomes in period 0 are by assumption never realized.
Finally, both $D_0$ and $D_1$ are everywhere observed.

A stronger, conditional version of stationarity identifies an average treatment effect on switchers in a \textit{fuzzy} design, where some units are initially treated so that $P(D_0 = 1) > 0$.

\begin{assumption}[Forward Mean Stationarity of Untreated Outcomes]
    \label{assn:fstat}
    The mean untreated outcome is constant across time, conditional on past treatment: 
    \[
        E [ Y_1 (0) - Y_0 (0) | D_0 = d_0] = 0 \quad \text{for $d_0 \in \{0,1\}$.}
    \]
\end{assumption}

\noindent
\autoref{assn:fstat} is jointly an assumption about stationarity (because it implies \autoref{assn:stat}) and about selection (because it conditions on treatment and implies \autoref{assn:fpbt}). 
Thus, it is similar to a mean version of conditional stationarity in \cite{rothsantanna}. 
However, it differs by conditioning on \emph{past} treatment in period 0, rather than treatment in period 1. 
As illustrated by the examples in \autoref{sec:examples} that violate parallel trends, conditioning on past treatment may be more credible in dynamic choice contexts where decision-makers are less likely to have (and thereby select on) information about the future than information about the past. 
In particular, under zero trend, \autoref{assn:fstat} is satisfied with learning in \autoref{thm:learning}, whereas its analog conditioning on treatment in period 1 is not.

Under \autoref{assn:fstat}, an analogous identification result is immediate upon applying \autoref{thm:id-tot} in the subpopulation where $D_0 = 0$. 

\begin{corollary}
    Under \autoref{assn:fstat}, the treatment effect in period 1 for the switchers into treatment is identified:
    \[
        E [ Y_{1} (1) - Y_{1} (0) | D_0 < D_1]
        = 
        \frac{E [ Y_1 | D_0 = 0] - E [ Y_0 | D_0 = 0]}{P (D_1 = 1| D_0 = 0)}.
    \]
\end{corollary}

\noindent
It is worth briefly comparing this to the weaker \autoref{assn:fpbt} result, which allowed for a non-zero conditional trend $\tau_0 \neq 0$. 
In that case, identification was not possible without further assumptions.

In the next subsection, we turn to the possibility of partial identification when such assumptions on trends are not viable.

\subsection{Bounds from Economic Structure}

We now consider partial identification of 
$
E [Y_1 (1) - Y_1 (0) | D_0 < D_1]
$
under assumptions stemming from our modeling of dynamic selection.%
\footnote{
For related but more model-agnostic approaches, see \cite{manskipepperdeterrence}    and \cite{rothrambachan}.
}
First, observe that the treated outcome $E [ Y_1 (1) | D_0 < D_1]$ is identified from the data, and so it suffices to focus on bounding the untreated outcome in period 1 among switchers into treatment, $E [ Y_1 (0) | D_0 < D_1]$. 
Alternatively, expand the treatment effect as: % in \eqref{eq:te-switchers-into},
\[
E [Y_1 (1) - Y_1 (0) | D_0 < D_1]
= 
E [Y_1 (1) - Y_0 (0) | D_0 < D_1]
-
E [Y_1 (0) - Y_0 (0) | D_0 < D_1].
\]
It suffices to bound the second term on the right, i.e., the switchers' untreated trend.

We can partially identify the average treatment effect on the switchers into treatment ($D_0<D_1$) using some version of monotone treatment on selection assumptions, similar to those  introduced by \cite{manski1997monotone} and \cite{manski2000monotone} but generalized to our multi-period setting.

\begin{assumption}[MTS]
{\,}
\label{assn:mts}
Monotone Treatment Selection (MTS) on  Level and Trend of Untreated Potential Outcomes:
\begin{enumerate}
    \item [a.] Level: $E [ Y_1 (0) | D_0 < D_1] \leq E [ Y_1 (0) | D_0 = D_1 = 0]$;
    \item [b.] Trend: $E [ Y_1 (0) - Y_0 (0) | D_0 < D_1] \geq E [ Y_1 (0) - Y_0 (0) | D_0 = D_1 = 0]$.
\end{enumerate}
\end{assumption}

\noindent
Although one could consider more general MTS assumptions across treatment sequences, we focus on these specific comparisons to the never-treated because they i) partially identify our target parameter and ii) have a basis in our model of learning about the control (\autoref{thm:learning-negative}).
The proof of the following identification result is immediate by the preceding discussion. 

\begin{proposition}
Under \autoref{assn:mts}, the identified set for the treatment effect $E[Y_1(1) - Y_1 (0) | D_0 < D_1]$ is:
\begin{align*}
\Big[
&
E [ Y_1  | D_1 > D_0] - E [Y_1  | D_0 = D_1 = 0],
\\
&E [ Y_1  - Y_0  | D_1 > D_0] 
- E [ Y_1  - Y_0  | D_0 = D_1 = 0]
\Big]
\end{align*}
\end{proposition}

We now motivate the identifying assumptions with the model of learning about the control (\autoref{thm:learning-negative}), where there is a pre-treatment period ($D_0=0$ for all units) and the parallel trends condition was typically violated. 
We begin by showing that, among decision-makers whose period 1 decision depends on the past realized outcome, those who continue without treatment are over-sampled from higher-mean untreated arms. 
This follows from two observations.
First, recalling \eqref{eq:cx-learn-signal}, past untreated outcomes provide further information about the true returns of the bandit arm, and thereby about future outcomes:%
\footnote{
Recall that $U_0$ summarizes the information at time 0, i.e.~$\mathcal{I}_0 = \{U_0\}$. In this section we use the $U_0$ notation to facilitate imposing additional structure.
}
\begin{equation}
    \label{eq:bound-u-1}
    E [ Y_1 (0) | U_0, Y_0 (0) = 1] \geq E [ Y_1 (0) | U_0, Y_0 (0) = 0].
\end{equation}
Second, for initial information sets 
$u^* \in \{ U_0 (\omega): \omega \in \Omega_{vl} \}$ 
of ``valuable learners'' in \eqref{eq:vl} where the period 1 treatment depends on the past outcome, we have the following equality (see \eqref{eq:D1-learning} in the appendix for details): 
\begin{equation}
    \label{eq:bound-u-2}
    \{ U_0 = u^*, D_0 = 0, D_1 = d_1\} = \{ U_0 = u^*, Y_0 (0) = 1 - d_1 \}  \quad \text{for $d_1 \in \{0,1\}$}
\end{equation}
Combining \eqref{eq:bound-u-1} and \eqref{eq:bound-u-2} leads to higher untreated outcomes $Y_1 (0)$ in period 1 among the never-treated than among those who switch into treatment:
\begin{equation}
\label{eq:bandit-persist-ineq-level}
E [ Y_1 (0) | U_0 = u^*, D_0 = D_1 = 0] \geq E [ Y_1 (0) |U_0 = u^*,  D_0 < D_1].
\end{equation}
Additionally, \eqref{eq:bound-u-2} implies that:
\begin{equation}
\label{eq:bound-u-3}
    E [ Y_0 (0) | U_0 = u^*, D_0 = 0, D_1 = d_1] = 1 - d_1
\end{equation}
and the fact that $Y_1 (0)$ was assumed binary implies:
\begin{equation}
\label{eq:bound-u-4}
    E [ Y_1 (0) | U_0 = u^*, D_0=0, D_1=d_1] \in [0,1]
\end{equation}
Combining \eqref{eq:bound-u-3} and \eqref{eq:bound-u-4} yields the opposite direction of inequality on the untreated trend: 
\begin{equation}
\label{eq:bandit-persist-ineq-trend}
E [ Y_1 (0) - Y_0(0)| U_0 = u^*, D_0 = D_1 = 0] 
\leq 0 \leq 
E [ Y_1 (0) - Y_0 (0) |U_0 = u^*,  D_0 < D_1].
\end{equation}
The directions of inequality in \eqref{eq:bandit-persist-ineq-level} and \eqref{eq:bandit-persist-ineq-trend} are those of \autoref{assn:mts}.
However, these inequalities also condition on a period 0 unobservable realization $U_0 = u^*$, for which the choice of period 1 treatment is a function of the period 0 realized outcome. 

To extend and aggregate the inequalities \eqref{eq:bandit-persist-ineq-level} and \eqref{eq:bandit-persist-ineq-trend} beyond valuable learners, one solution is to impose (or ideally, derive) additional structure between treatment sequences, untreated outcomes, and unobservables.
For example, consider the following assumption to rationalize MTS on the level of untreated potential outcomes (\autoref{assn:mts}a).

\begin{assumption}
\label{assn:mts-u}
The initial information  $U_0 \in \mathbb{R}$ satisfies: 
\begin{enumerate}
    \item[a.] For all $u$ where the conditional expectations exist, 
    \begin{align*}
        E [ Y_1 (0) | U_0 = u, D_0 < D_1] 
        &\leq
        E [ Y_1 (0) | U_0 = u, D_0 = 0] 
        % E [Y_1 (0) | U_0 = u, D_0 = D_1 = 0] 
        % &\geq 
        % E [ Y_1 (0) | U_0 = u, D_0 = 0].
    \end{align*}
    \item[b.] $E [ Y_1 (0) | U_0 = u, D_0 = 0]$ is non-decreasing in $u$.
    \item[c.] $(U_0 | D_0 < D_1) \preceq_{FOSD} (U_0 | D_0 = D_1 = 0)$ 

\end{enumerate}
\end{assumption}

\noindent
\autoref{assn:mts-u}a generalizes \eqref{eq:bandit-persist-ineq-level} across all unobservables $U_0 = u$: 
selection into treatment in period 1 is indicative of lower untreated outcomes in expectation.
\autoref{assn:mts-u}b imposes that higher unobservables correspond to higher untreated outcomes; for example, the unobservable could be an average return from a previous, fixed-length sequence of untreated outcome realizations observed by decision-makers but not the econometrician. 
Note that this has no content on its own, since unobservables $U_0$ can always be relabeled and arranged such that this condition is satisfied. 
However, \autoref{assn:mts-u}c imposes that unconditional selection into treatment oversamples lower unobservables, which by \autoref{assn:mts-u}b correspond to lower mean expected returns.
Under \autoref{assn:mts-u}, we recover \autoref{assn:mts}a as follows:
\begin{align*}
E [ Y_1 (0) | D_0 < D_1] 
&= 
E [ E [ Y_1 (0) | U_0, D_0 < D_1] | D_0 < D_1] \\
&\leq
E [ E [ Y_1 (0) | U_0, D_0 = 0] | D_0 < D_1] \\
&\leq
E [ E [ Y_1 (0) | U_0, D_0 = 0] | D_0 = D_1 = 0] \\
&\leq
E [ E [ Y_1 (0) | U_0, D_0 = D_1 = 0] | D_0 = D_1 = 0] \\
&= 
E [ Y_1 (0) | D_0 = D_1 = 0] 
\end{align*}
where the equalities follow by the law of iterated expectations, the first and third inequalities follow from \autoref{assn:mts-u}a, and the second inequality follows from combining \autoref{assn:mts-u}b and c.
Alternatively, one could attempt to (partially) identify a marginal group $U_0 = u^*$ with other assumptions or an additional source of exogenous variation, such as an instrument.
We leave this to future work.

\section{Conclusion}
\label{sec:conclusion} 

In this paper, we made connections between the commonly used parallel trends assumption and models of dynamic rational choice in economics. In particular, we highlight models with time-varying treatment costs, learning, correlated utilities, and optimal stopping. 
Our aim is to focus on channels of dynamic behavior that are at work in these models, in order to understand the way that these channels can validate or invalidate the design-based identifying assumption of parallel trends. 
The examples we provide are deliberately stylized and simple. 
In cases when parallel trends may be violated, we provide pointers to inference approaches based on simple and familiar economic restrictions that are motivated by economic concerns.  
These include either relaxing or considering alternatives to parallel trends. 

Our hope is that the paper provides a canvas by which further work on the econometrics of treatment or causal inference with observational data can be examined. 
This is particularly important with dynamic decisions in rich environments where a variety of preference- and information-based dynamic considerations
play a role and where it is helpful to relate choice models based on these considerations to examine what behavior is allowed and what is not under the parallel trends condition.
We conclude with a sentiment consistent with our approach in this paper, namely that ``models are most useful when they are used to challenge existing formulations, rather than to validate or verify them'' (\cite{oreskes1994verification} p. 644).    
In that sense, using choice models to shed light on DiD regressions is a common and worthy use of modeling. 

\newpage 
\appendix

\section{Proofs}
\label{apx:proofs}

\begin{proof}[Proof of \autoref{thm:pt-dc}] 
    The fact that the parallel trends condition implies the set of pairwise equalities \eqref{eq:pt-dc1} and \eqref{eq:pt-dc2} is immediate. 
    Conversely, suppose \eqref{eq:pt-dc1} holds and define $\tau = E [ Y_1 (0) - Y_0 (0) | D_0 = d_0]$, which is constant and does not vary with $d_0$.
    It suffices to show that \eqref{eq:pt-dc1} and \eqref{eq:pt-dc2} imply $E [ Y_1 (0) - Y_0 (0) | D_0 = d_0, D_1 = d_1] = \tau$ for each $(d_0,d_1)$.
    Define $p(d_0) = E [ D_1 | D_0 = d_0]$. Then for each $t$ the law of total probability implies: 
    \begin{align*}
        E [Y_t (0) | D_0 = d_0] 
        =& p(d_0) E [Y_t (0) | D_0 = d_0, D_1 = 1] \\
        +& (1-p(d_0)) E [ Y_t (0) | D_0 = d_0, D_1 = 0]
    \end{align*}
    and thus 
        \begin{align*}
        E [Y_1 (0) - Y_0 (0) | D_0 = d_0] 
        =& p(d_0) E [Y_1 (0) - Y_0 (0) | D_0 = d_0, D_1 = 1] \\
        +& (1-p(d_0)) E [ Y_1 (0) - Y_0 (0) | D_0 = d_0, D_1 = 0].
    \end{align*}
By \eqref{eq:pt-dc2} both differences on the right are equal to the term on the left, which in turn is equal to $\tau$.
\end{proof}

\begin{proof}[Proof of the claim in \autoref{ex:violate-bs}]
    We first show that $Y_0 (0) = Y_1 (0)$ a.s.~is necessary for parallel trends. 
    First, observe that $D_0 = 0$, $D_1 = 1-Y_0$ and $P(D_1 = 1) \in (0,1)$ imply: 
    \begin{equation}
    \label{eq:ex1-y0}
        E [ Y_0 (0) | D_1 = d] = 1-d.
    \end{equation}
     Substituting into the deviation from parallel trends,
    \begin{align*}
        &
        E [ Y_1 (0) - Y_0 (0) | D_1 = 1] - E [ Y_1 (0) - Y_0 (0) | D_1 = 0] \\
        =& 
        E [ Y_1 (0) | D_1 = 1] - E [ Y_1 (0) | D_1 = 0] + 1,
    \end{align*}
    Because outcomes $Y_1 (0)$ are assumed binary, this deviation is zero only if:  
    \[
        E [ Y_1 (0) | D_1 = d] = 1-d \quad \text{for $d=0,1$.}
    \]
    Combining with \eqref{eq:ex1-y0}, this in turn requires $Y_0 (0) = Y_1 (0)$ almost surely because the potential outcomes are binary.
    Conversely, $Y_0 (0) = Y_1 (0)$ a.s.~is sufficient for parallel trends. 
\end{proof}

\begin{proof}[Proof of the claims in \autoref{thm:persistent}]
In period 0, the treatment rule for $D_0$ is a function of initial information $U_0$ alone by \eqref{eq:opt_D0}. 
Furthermore, under \eqref{eq:info-persist}, the treatment rule at $t=1$ is:
\begin{align}
        D_1 (d_0) 
        &= \mathbf{1} \{ E [ Y_1 (1) - Y_1 (0) - (K_1 (d_0,1) - K_1 (d_0, 0)) | U_0, U_1(d_0) ] \geq 0 \} \notag \\
        &= \mathbf{1} \{ E [ Y_1 (1) - Y_1 (0) - (K_1 (d_0,1) - K_1 (d_0, 0)) | U_0 ] \geq 0 \}. \label{eq:D1-I0}
\end{align}
Hence in period 1, the treatment rule for $D_1 (d_0)$ is also a function of $U_0$ alone. 
It then follows that the realized treatment at $t=1$, i.e., $D_1 = D_1(D_0)$, is also a function of $U_0$ alone. 
By the law of iterated expectations and \eqref{eq:info-persist-pt}, we have:
\begin{align*}
    E [ Y_1 (0) - Y_0 (0) | D_0 = d_0, D_1 = d_1] 
     = E [ E [ Y_1 (0) - Y_0 (0) | U_0 ] | D_0 = d_0, D_1 = d_1 ] = \tau
\end{align*}
for all $d_0, d_1 \in \{0,1\}$.
\end{proof}

\begin{proof}[Proof of the claims in \autoref{thm:learning}]
% t = 0
The optimal treatment rule \eqref{eq:opt_D0} in period 0:
\begin{equation}
    \label{eq:d0-bandit}
    D_0 = \mathbf{1} \{ E [V_0(1) - V_0 (0) + \beta (W_1 (1) - W_1 (0)) | U_0] \geq 0 \} 
\end{equation}
is a function of $U_0$ alone. 
By the law of iterated expectations and \eqref{eq:info-persist-pt},
\[
    E [ Y_1 (0) - Y_0 (0) | D_0 = d_0] 
    = E [ E [ Y_1 (0) - Y_0 (0) | U_0] | D_0 = d_0] 
    = \tau. 
\]
With no treatment in period 0 and because $U_1 (d_0) = Y_0 (d_0)$, the optimal treatment rule \eqref{eq:opt_D1} in period 1 is: 
\begin{align*}
    D_1 (0) 
    &=
    \mathbf{1} \{ E [ Y_1 (1) - Y_1 (0) - (K_1 (0,1) - K_1 (0, 0)) | U_0, Y_0 (0) ] \geq 0 \} \\
    &=
    \mathbf{1} \{ E [ Y_1 (1) - Y_1 (0) - (K_1 (0,1) - K_1 (0, 0)) | U_0 ] \geq 0 \} 
\end{align*}
where the second equality holds because of \eqref{eq:control-no-learn} and because costs are known initially $K \in \mathcal I_0$.
With $D_1 (0)$ and $D_0$ both being functions of $U_0$ alone, we can apply the law of iterated expectations and use the condition \eqref{eq:info-persist-pt} to show:
\begin{align*}
    E [ Y_1 (0) - Y_0 (0) | D_0 = 0, D_1 = d_1] 
    &= 
    E [ Y_1 (0) - Y_0 (0) |D_0 = 0, D_1 (0) = d_1 ] \\
    &= 
    E [ E [ Y_1 (0) - Y_0 (0) | U_0 ] | D_0 = 0, D_1 (0) = d_1 ] = \tau
\end{align*}
for all $d_1 \in \{ 0,1 \}$.

With treatment in period 0, the optimal treatment rule \eqref{eq:opt_D1} in period 1 is: 
\[
    D_1 (1) = 
    \mathbf{1} \{ E [ Y_1 (1) - Y_1 (0) - (K_1 (0,1) - K_1 (0, 0)) | U_0, Y_0 (1) ] \geq 0. \} %\label{eq:d11}
\]
Again, by the law of iterated expectations: 
\begin{align*}
    E[Y_1 (0) - Y_0 (0) | D_0 = 1, D_1 = d_1] 
    &= 
    E[Y_1 (0) - Y_0 (0) | D_0 = 1, D_1 (1) = d_1] \\
    &=
    E[ E [ Y_1 (0) - Y_0 (0) | U_0, Y_0 (1)] | D_0 = 1, D_1 (1) = d_1]. 
\end{align*}
Note:
\[
    E [ Y_1 (0) - Y_0 (0) | U_0, Y_0(1)] = E [ Y_1 (0) - Y_0 (0) | U_0 ] = \tau,
\]
where the first equality is due to \eqref{eq:control-no-learn} and \eqref{eq:no-feedback-safe}, and the second due to \eqref{eq:info-persist-pt}.
Hence
\[
    E[Y_1 (0) - Y_0 (0) | D_0 = 1, D_1 = d_1] = \tau
\]
for $d_1 \in \{0,1\}$.
It then follows from \autoref{thm:pt-dc} that the parallel trends condition holds.
\end{proof}

\begin{proof}[Proof of the claims in \autoref{thm:learning-negative}]
With no one treated initially ($D_0 = 0$), and costs known to decision-makers ($K\in \mathcal I_0$), the optimal treatment rule \eqref{eq:opt_D1} in period 1 becomes: 
\begin{equation}
    \label{eq:D1-learning}
    D_1 = D_1 (0) = \mathbf{1} \{  E [ Y_1 (1) | U_0] - E [ Y_1 (0) | U_0, Y_0 (0)] \geq \tilde{K}_1 \}
\end{equation}
because there is no learning across arms in \eqref{eq:no-learning-across}. 

Combining the decision rule \eqref{eq:D1-learning} and the assumption on learning in \eqref{eq:cx-learn-signal}, we can partition the sample space into three groups based on the response to (possibly counterfactual) information from the past realized outcome $Y_0 (0)$: those who are always ($a$) or never ($n$) treated in period 1 regardless of $Y_0 (0)$,
\begin{eqnarray*}
    \Omega_{a} 
    &\equiv &
    \{ \omega \in \Omega: E[Y_{1}(1) -\tilde{K}_{1} |U_{0}(\omega)]\geq E[Y_{1}(0)|U_{0} (\omega),Y_{0}(0)=1]\} \text{;} \\
    \Omega_{n} 
    &\equiv& 
    \{ \omega \in \Omega: E[Y_{1}(1) -\tilde{K}_{1} |U_{0} (\omega)] <E[Y_{1}(0)|U_{0} (\omega),Y_{0}(0)=0]\};
\end{eqnarray*}%
and those for whom $D_1(0)$ depends on $Y_0 (0)$ because there is valuable learning ($vl$), which was defined in \eqref{eq:vl} and restated here for convenience:%
\footnote{
    For the always- and never-treated, learning from past outcome does not affect their future decisions, and hence is not considered ``valuable''.}  
\[
    \Omega _{vl} \equiv 
    \{ \omega \in \Omega: E[Y_{1}(0)|U_{0} (\omega),Y_{0}(0)=0]\leq
    E[Y_{1}(1) -\tilde{K}_{1} |U_{0} (\omega)]<E[Y_{1}(0)|U_{0} (\omega),Y_{0}(0)=1]\}.
\]
Even though the partition is defined based on the responses to period-0 outcomes, it is by construction measurable with respect to initial information $U_0$. 
Also, the groups are defined in terms of expectations conditional on counterfactual period-0 outcomes -- that is, for values of $Y_0 (0)$ that may differ from those realized.%
\footnote{
    For simplicity of notation we assume the expectations are well-defined. This is so except where potential outcomes $Y_0 (0)$ are degenerate given $U_0$. 
    At such values of initial information there is no rational learning, in which case the ill-defined history-conditional expectations $E[Y_1 (0) | U_0, Y_0 (0)]$ can be replaced with the well-defined history-unconditional expectations $E [ Y_1 (0) | U_0]$.}
We can further partition the ``valuable learning'' group based on their \textit{realized} period-0 outcomes: 
\begin{eqnarray*}
    \Omega_{vl}^0 &=& \{ \omega \in \Omega_{vl}:  Y_0(0;\omega) = 0 \},
    \\
    \Omega_{vl}^1 &=& \{ \omega \in \Omega_{vl}: Y_0(0;\omega) = 1 \}.
\end{eqnarray*}
Let 
$P_i^j \equiv P ( \Omega_i^j )$ for $i \in \{a, n, vl\}$ and $j \in \{0,1\}$. 

We can then write
\begin{align*}
    &E [ Y_1 (0) - Y_0 (0) | D_1 = 1, D_0 = 0 ] 
    =
    E [ Y_1 (0) - Y_0 (0) | \Omega_a \cup \Omega_{vl}^0 ] \\
    =& 
    \frac{P_a}{P_a + P_{vl}^0} E [ Y_1 (0) - Y_0 (0) | \Omega_a]
    + 
    \frac{P_{vl}^0}{P_a + P_{vl}^0} E [ Y_1 (0) - Y_0 (0) | \Omega_{vl}^0] \\
    =& 
    \frac{P_a}{P_a + P_{vl}^0} \, \tau
    + 
    \frac{P_{vl}^0}{P_a + P_{vl}^0} E [ Y_1 (0) | \Omega_{vl}^0],
\end{align*}
where the last equality follows from the fact that $\Omega_a$ only restricts the initial information $U_0(\omega)$, combined with an application of the law of iterated expectation to
(\ref{eq:info-persist-pt}), and from the fact that $Y_0 (0) = 0$ among $\Omega_{vl}^0$.
By analogous reasoning, 
\begin{align*}
    &E [ Y_1 (0) - Y_0 (0) | D_1 = 0, D_0 = 0 ] 
    =
    E [ Y_1 (0) - Y_0 (0) | \Omega_n \cup \Omega_{vl}^1 ] \\
    =& 
    \frac{P_n}{P_n + P_{vl}^1} E [ Y_1 (0) - Y_0 (0) | \Omega_n]
    + 
    \frac{P_{vl}^1}{P_n + P_{vl}^1} E [ Y_1 (0) - Y_0 (0) | \Omega_{vl}^1] \\
    =& 
    \frac{P_n}{P_n + P_{vl}^1} \, \tau 
    + 
    \frac{P_{vl}^1}{P_n + P_{vl}^1} ( E [ Y_1 (0) | \Omega_{vl}^1] - 1 ). 
\end{align*}

Parallel trends hold if and only if both preceding equations are equal to the same constant. Because $E[Y_1(0)-Y_0(0)]=\tau$ by (\ref{eq:info-persist-pt}), that constant must be $\tau$.
This latter condition (that both preceding equations are equal to the constant $\tau$) holds under either one of the following two sufficient conditions:
either (i) ``there is zero probability of valuable learning, $P_{vl} = P_{vl}^0 + P_{vl}^1 = 0$", or (ii) ``valuable learning occurs where untreated outcomes are identical over time almost surely, $P(Y_0 (0) = Y_1 (0) | \Omega_{vl}) = 1$, and the trend is zero ($\tau=0)$".%
\footnote{
    This still allows for learning about the untreated arm.
    For example, there is initial uncertainty about whether the process is one of two degenerate arms, so that $E [Y_1 (0) | \Omega_{vl}] \in (0,1)$.
    However, once the first return is observed, this uncertainty is fully resolved, i.e., $E[Y_1 (0) | \Omega_{vl}, Y_0(0)] \in \{0,1\}$.}

The sufficiency of the first condition (i) is straightforward. 
For the sufficiency of the second condition (ii), suppose $P_{vl} > 0$, which in turn can be split into two cases. 
First, if $P_{vl}^0 > 0, P_{vl}^1 > 0$, then parallel trends requires that:
\[
    E [ Y_1 (0) | \Omega_{vl}^0]  = E [ Y_1 (0) | \Omega_{vl}^1] - 1 = \tau
\]
Because $Y_1 (0)$ is binary, $E [ Y_1 (0) | \Omega_{vl}^0] \geq 0$ while $ E [Y_1 (0) | \Omega_{vl}^1] -1 \leq 0$. So, the two equalities above are only possible together if $\tau = 0$, which happens if $P(Y_0 (0) = Y_1(0) | \Omega_{vl}) = 1$, and there is a zero trend, $\tau=0$. 
Sufficiency of condition (ii) when either $P_{vl}^0$ or $P_{vl}^1 = 0$ is zero but not both follows from an analogous argument, and is omitted for brevity.
\end{proof}

\begin{proof}[Proof of the claims in \autoref{ex:roy}]
    That conditions \ref{item:stat-roy} and \ref{item:as-roy} are sufficient for the parallel trends condition is immediate from the fact they jointly imply: 
    \[
        E[ Y_1 (0) - Y_0 (0) | D_0 = d_0, D_1 = d_1 ] = 0 
    \]
    for all $(d_0, d_1)$ occurring with positive probability.
    Conversely, suppose the Roy model holds, so that:
    \begin{equation}
        \label{eq:roy-imply}
                D_t = 0 \implies Y_t (0) = 1 .
    \end{equation}
    We consider two cases based on whether there is an interior probability of never-treated observations, $P (D_0 = D_1 = 0) > 0$.
    First, if there exist never-treated, then \eqref{eq:roy-imply} implies:
    \[
        E [ Y_1 (0) - Y_0 (0) | D_0 = D_1 = 0] = 1 - 1 = 0
    \]
    Therefore the parallel trends condition requires stationarity (Condition 1).
    Again invoking \eqref{eq:roy-imply}, parallel trends also requires that:
    \[
        E [ Y_{1-t} (0) | D_t =0, D_{1-t}=1] = 1 \quad \text{for $t \in \{0,1\}$}
    \]
    Aggregating over the set of ever-untreated $D_0 D_1 = 0$ yields Condition \ref{item:as-roy}.

    Second, if there do not exist never-treated, then by assumption that $P (D_t = 1) = P(Y_t (1) \geq Y_t (0)) \in (0,1)$ for $t=0,1$, there must exist movers into and out of treatment each period, namely $P(D_0 > D_1), P(D_0 < D_1) \in (0,1)$. 
    Then \eqref{eq:roy-imply} implies:
    \begin{align*}
        E [ Y_0 (0) | D_0 < D_1] &= 1 \\
        E [ Y_1 (0) | D_0 > D_1] &= 1.
    \end{align*}
    In turn, binary outcomes then imply that: 
    \[
        E [ Y_1 (0) - Y_0 (0) | D_0 < D_1] \leq 0 \leq E [ Y_1 (0) - Y_0 (0) | D_0 > D_1]
    \]
    Therefore the parallel trends condition requires stationarity (Condition 1). Combined with the preceding restrictions on degenerate outcomes, this also requires Condition \ref{item:as-roy}. 
\end{proof}

\begin{proof}[Proof of \autoref{thm:ind-exogeneity}]
    Sufficiency of strict exogeneity \eqref{eq:strict-exogeneity} for parallel trends is immediate since then: 
    \[
        E [ Y_1 (0) - Y_0 (0) | D_0 = d_0, D_1 = d_1] = \mu_1 - \mu_0
    \]
    for constants $\mu_0, \mu_1 \in \mathbb{R}$ and all $d_0, d_1 \in \{0,1\}$.
    Conversely, suppose treatment decisions satisfy \eqref{eq:selection-present} and $(U_t, Y_t (\cdot))$ satisfy \eqref{eq:uy-ind}. 
    Since treatments $D_t$ are a function of $U_t$ by \eqref{eq:selection-present}, it follows from \eqref{eq:uy-ind} that
    $D_t \perp (Y_{1-t} (0), D_{1-t})$, and therefore $D_t \perp Y_{1-t} (0) | D_{1-t}$, for $t \in \{0,1\}$. In that case, we have:
    \begin{equation}
    \label{eq:ind-t1} 
        E [ Y_t (0) | D_0 = d_0, D_1 = d_1] = E [ Y_t (0) | D_t = d_t].
    \end{equation}
    (Note that this is testable evaluating at $d_t = 0$ and comparing across $d_{1-t} = 0,1$.)
    
    If $E[D_t] \in \{0,1\}$ for each $t \in \{0,1\}$, then the implication \eqref{eq:strict-exogeneity} is trivial. 
    Therefore suppose $E [D_0] \in (0,1)$. Then for a realized period 1 outcome occurring with positive probability, say $D_1 = 0$, the parallel trends condition combined with \eqref{eq:ind-t1} requires:
    \[
        E [ Y_1 (0) | D_1 = 0] - E [ Y_0 (0) | D_0 = 0] = E [ Y_1 (0) | D_1 = 0] - E [ Y_0 (0) | D_0 = 1]
    \]
    or $E [ Y_0 (0) | D_0 = 0] = E [ Y_0 (0) | D_0 = 1] = \mu_0$ for some constant $\mu_0 \in \mathbb{R}$.
    Invoking again \eqref{eq:ind-t1}, it follows that:
    \[
        E [ Y_0 (0) | D_0 = d_0, D_1 = d_1] = \mu_0
    \]
    for all realizations $(d_0, d_1)$ occurring with positive probability. An analogous argument holds to show \eqref{eq:strict-exogeneity} for $t=1$.
\end{proof}

\begin{proof}[Proof of \autoref{thm:strong-roy-os}]
As in the proof of \autoref{ex:roy}, that conditions \ref{item:stat-roy} and \ref{item:as-roy} are sufficient for the parallel trends condition is immediate.
Conversely, suppose the Roy model with irreversible treatment (henceforth the ``constrained'' model) holds. 
Then the optimal treatment rules \eqref{eq:opt_D1} and \eqref{eq:opt_D0} reduce to:
\begin{align}
    D_1 (0) &= \mathbf{1} \{ Y_1 (1) - Y_1 (0)  \geq 0 \} \label{eq:strong-roy-os1} \\
    D_0 &= \mathbf{1} \{ Y_0 (1) - Y_0 (0) + \beta \min \{ Y_1 (1) - Y_1 (0), 0 \} \geq
    0 \}.  \label{eq:strong-roy-os0}
\end{align}
To connect to the logic of the Roy model with $K_t(\mathbf{d}^t)=0$ (henceforth the ``unconstrained'' model), it is useful to recall and separately define the (counterfactual) unconstrained choices,  $\tilde{D}_t \equiv \mathbf{1} \{ Y_t (1) \geq Y_t (0) \}$.
By definition, in period 1 we have $D_1 (0) = \tilde{D}_1 $;%
\footnote{
It is useful to note, however, that we may have $D_1 \neq \tilde{D}_1$ when the constraint binds and $D_0 = 1$.
}
in period 0 we have $D_0 \leq \tilde{D}_0$, since the lost option value from treatment in period 1 weakly discourages treatment in period 0, i.e., $\beta \min \{ Y_1 (1) - Y_1 (0) , 0\} \leq 0$.
If $D_1 (0) = 1$, then $Y_1 (1) \geq Y_1 (0)$, and so this option value is 0 and $D_0 = \tilde{D}_0$.
It follows from $D_1 (0) = \tilde{D}_1 $, $D_0 \leq \tilde{D}_0$, and $D_1 (0) = 1 \implies D_0 = \tilde{D}_0$ that the constrained and unconstrained movers into treatment are the same: 
\[
    \{ D_0 < D_1 \} = \{ \tilde{D}_0 < \tilde{D}_1  \}. 
\]
From \eqref{eq:roy-imply}, it follows that among the constrained movers into treatment we also have $Y_0 (0) = 1$. 
The constrained never-treated consist of i) the unconstrained never-treated and ii) the subset of the unconstrained movers out of treatment $\{ \tilde{D}_0 > \tilde{D}_1 \}$ who change their choice in period 0 because of the constraint:%
\footnote{
    The unconstrained movers out of treatment who change their decision in period 0 because of the constraint must have a strictly positive continuation value, which implies $Y_1 (0) > Y_1 (1)$. Furthermore, the choice rule \eqref{eq:strong-roy-os1} and $\beta \leq 1$ imply $Y_0 (1) = Y_0 (0)$. 
    Thus, the subset of unconstrained movers out of treatment impacted by the constraint are intuitively those with a relatively high option value of waiting.
}
\[
    \{ D_0 = D_1 = 0 \} = \{ \tilde{D}_0 = \tilde{D}_1 = 0 \} \, \cup \, \{ \tilde{D}_0 > D_0 = \tilde{D}_1 \}.
\]
Again from \eqref{eq:roy-imply}, it follows that among the constrained never-treated we have $Y_1 (0) = 1$. 

By the logic of the unconstrained strong Roy model (\autoref{ex:roy}), all unconstrained movers into treatment have a weakly negative untreated trend, all unconstrained never-treated have a zero untreated trend, and all unconstrained movers out of treatment have a weakly positive untreated trend:
\begin{align*}
    \tilde{D}_0 < \tilde{D}_1 & \implies  Y_1 (0) - Y_0 (0) \leq 0  \\ %
    \tilde{D}_0 = \tilde{D}_1 = 0 &\implies Y_1 (0) - Y_0 (0)   = 0 \\ %  
    \tilde{D}_0 > \tilde{D}_1 & \implies   Y_1 (0) - Y_0 (0)    \geq 0  % 
\end{align*}
By assumption that constrained movers into treatment and constrained never-treated each exist with positive probability, it follows that the groups have weakly negative and positive untreated trends, respectively: 
\begin{align*}
    E[ Y_1 (0) - Y_0 (0) | D_0 < D_1] \leq 0 \leq E[ Y_1 (0) - Y_0 (0) | D_0 = D_1 = 0]
\end{align*}
Thus, the parallel trends condition requires stationarity (Condition 1). The necessity of Condition 2 then follows because $Y_0 (0) = 1$ among movers into treatment and $Y_1 (0) = 1$ among the constrained never-treated. 
\end{proof}

\begin{proof}[Proof of \autoref{thm:id-tot}]

We have:
\begin{align*}
&E[Y_{1}(1)-Y_{1}(0)\mid D_{0}<D_{1}] \\
& \stackrel{1}{=} E[Y_{1}(1)-Y_{1}(0)\mid D_{1}=1] \\
& \stackrel{2}{=} E[Y_{1}\mid D_{1}=1]-E[Y_{1}(0)\mid D_{1}=1] \\[0.1in]
& \stackrel{3}{=} \frac{E[Y_{1}\mid D_{1}=1] P (D_{1}=1)-E[Y_{1}(0)\mid D_{1}=1] P
(D_{1}=1)}{\Pr (D_{1}=1)} \\[0.1in]
& \stackrel{4}{=} \frac{E[Y_{1}\mid D_{1}=1]P (D_{1}=1)-\{E[Y_{1}(0)]-E[Y_{1}(0)\mid
D_{1}=0]P (D_{1}=0)\}}{P (D_{1}=1)} \\[0.1in]
& \stackrel{5}{=} \frac{E[Y_{1}\mid D_{1}=1]P (D_{1}=1)-E[Y_{0}(0)]+E[Y_{1}\mid
D_{1}=0]P (D_{1}=0)}{P (D_{1}=1)} \\
& \stackrel{6}{=} \frac{E(Y_{1})-E(Y_{0})}{P (D_{1}=1)} 
\end{align*}
The first equality follows because no one is treated in period 0, i.e.~$D_{0}=0$ almost surely.
The fourth equality follows from the law of iterated expectation.
The fifth equality follows from \autoref{assn:stat}. 
The sixth equality follows from the law of iterated expectation and the fact that $D_{0}=0$ almost surely.
\end{proof}

\section{Selection and Parallel Trends: Heuristics}
\label{apx:heuristics}

As discussed in Section 3 and 4, a dynamic utility maximization (DUM) model allows individual decision makers to self-select into treatment sequences based on their information sets. 
Such endogenous self-selection could invalidate the parallel trends (PT) condition, especially when the information sets are correlated with potential outcomes. 
In this part of the appendix, we provide a list of heuristic guidelines, which may help empiricists to evaluate the plausibility of parallel trends (PT) in DiD analyses. 
This list is motivated by our results only, and therefore is by no means exhaustive. 

\begin{enumerate}

    \item {\it Direct selection on past outcomes.} 
        As in \cite{ashenfelter1978}, PT is suspect if the treatment decision is a function of past outcome. (Example 1)  
        
    \item {\it Indirect selection on past outcomes through learning.} 
        The presence of imperfect information about outcomes at the time of treatment, and thus the possibility of past outcome realizations affecting present treatment through learning, can potentially invalidate PT.   
        In particular, PT is suspect if there is learning about \textit{untreated} (potential) outcomes. (Examples 3 and 4)
        
        On the other hand,  {\it forward-looking experimentation}  need not invalidate (partial) PT necessarily since the treatment decision depends on outcome realizations that DM do not yet know. (Example 4) 
        
    \item {\it Roy-style selection on present outcomes.} 
        If the treatment in each period is a rational choice based on factors that are correlated with the potential outcomes, and if these factors and outcomes are insufficiently correlated over time, then PT is likely to be violated due to selection. (Example 5) 
        
    \item {\it Staggered designs and irreversible treatments.} 
        These introduce inevitable forward-looking considerations based on the foregone option value of delaying the irreversible choice until next period.  
        Similar to the case of experimentation, such considerations may change the margin of treatment without violating parallel trends per se. 
        However, previous concerns about learning and selection on present outcomes remain. (Example 6 and Section 4.4) 
    
    \item {\it Anticipation and latent choices.} 
        As in \cite{malani2015anticipation}, if the expectation of future treatments affects present \textit{potential} outcomes then this can lead to violations of PT.\footnote{
        Again, note that the expectation of future treatments can affect present realized outcomes through treatment decisions with forward-looking motives, without necessarily violating parallel trends.} 
        More generally, if potential outcomes are indexed by other latent choices, these may lead to violations of PT. (Section 4.5)
\end{enumerate}

\section{Illustration with Ashenfelter and Card (1985)}
\label{apx:illustrate}

We illustrate by linking our overall framework based on DUM and the relevant examples of Section 4 to the simplified setting of the job training application in \cite{ashenfeltercard1985training} [AC1985]. 
There, the sample consist of individuals $i=1,2,...,N$ and two periods $t\in \{0,1\}$. 
At time $t$, each individual $i$  receives a binary treatment $D_{it}\in \{0,1\}$, with $D_{it}=1$ if $i$ joints the job training program at time $t$. 
In the examples below, we may let $Y_{it}(\cdot )$ be either discrete (e.g. employment status for $i$ at time $t$) or continuous (e.g., wage compensation for $i$ at time $t$). 
In this case, our Assumption 0 posits an individual $i$'s potential outcome $Y_{it}(d_{it})$ depends only on $i$'s own contemporaneous treatment $d_{it}$.\medskip 

\noindent \textit{Example 1. (Selection on past outcomes.)} 
Suppose the potential outcome of interest is an individual $i$'s employment status at time $t$, i.e., $Y_{it}(0),Y_{it}(1)\in \{0,1\}$. 
Let there be a pre-treatment period, i.e., no one joins the job training program initially so that $D_{i0}=0$ for all $i$. Also, suppose $D_{i1}=1-Y_{i0}$. 
That is, only those unemployed at $t=0$ choose to join the job training program at $t=1$. 
In this case, the parallel trends condition holds if and only if the potential employment status of $i$ without participating in the training program does not change over time $t=0,1$. 
That is, $Y_{i0}(0)=Y_{i1}(0)$ for all $i$. 

To set up the stage for further examples, we fit the AC1985 application
within our unifying framework, under which the sequence of treatments are
determined endogenously by dynamic utility maximization (DUM). We suppress
individual subscripts $i$ from notation in what follows. Let the initial
information $\mathcal{I}_{0}=\{U_{0}\}$ subsume (time-invariant)\ individual
heterogeneity --- say, $\omega _{i}$ in equation (1) in AC1985 --- as well
as the discount factor $\beta $ and pre-determined fees for program
participation, 
\begin{equation}
K_{t}(0)=0\text{ and }K_{t}(1)=k\text{ for }t=0,1\text{.}  \label{eq:costs}
\end{equation}
Let (potential) incremental information $U_{1}(d_{0})$ consist solely of
realized past outcomes, i.e., $\mathcal{I}_{1}(d_{0})=\{U_{0},U_{1}(d_{0})\}$
with $U_{1}(d_{0})=Y_{0}(d_{0})$. Let the net (static)\ payoff at time $t$
only depend on contemporary treatment status, i.e., $%
V_{t}(d_{it})=Y_{it}(d_{it})-K_{t}(d_{it})$. Then the decision rules in (7)
and (8) are simplified to%
\begin{eqnarray}
D_{1}(d_{0}) &=&\mathbf{1}\{E\left[ Y_{1}(1)-Y_{1}(0)\mid U_{0},Y_{0}(d_{0})\right]
\geq k\}\text{,}  \label{eq:D1} \\
D_{0} &=&\mathbf{1}\{E\left[ Y_{0}(1)-Y_{0}(0)+\beta \left( W_{1}(1)-W_{1}(0)\right)
\mid U_{0}\right] \geq k\}\text{,}  \label{eq:D0}
\end{eqnarray}%
where 
\[
W_{1}(d_{0})=\max \{\text{ \ }E\left[ Y_{1}(1)\mid U_{0},Y_{0}(d_{0})\right]
-k\text{ \ },\text{ \ }E\left[ Y_{1}(0)\mid U_{0},Y_{0}(d_{0})\right] \text{
\ }\}
\]
is the optimal continuation value.\medskip 

\ \ 

\noindent \textit{Example 2. (No learning about potential outcomes.)}
Suppose the potential outcomes $Y_{t}(d_{t})$ are wage compensations, and the sequence of decisions to join the job training program are made by rational, forward-looking individuals as under (DUM). Assume an individual's heterogeneity in the initial information set does not affect the average trend in his wage without any training, as stated in (11).
Furthermore, assume that once conditional on such individual heterogeneity, the past  wages $Y_{0}(d_{0})$ are never informative about average wages $Y_{1}(d_{1})$ in the future, regardless of how the potential treatment status varies over time (i.e., regardless of whether $d_{0}=d_{1}$ or not), as stated in (12). 
Then the parallel trends condition holds. 

\ \ 

\noindent \textit{Example 3. (Learning only under treatment.)} 
As in Example 2, suppose the potential outcomes $Y_{t}(d_{t})$ are wage compensations, and the sequence of decisions to join the job training program are made by rational, forward-looking individuals as under (DUM). 
Suppose an individual's heterogeneity in the initial information set does not affect the average trend in his wage without any training. 
In addition, suppose that once conditional on individual heterogeneity in $U_{0}$, an individual can only learn about his expected wage with training in the future $Y_{1}(1)$ if he observes past wages with training $Y_{0}(1)$; no other scenarios of learning beyond the influence of initial heterogeneity $U_{0}$ is possible.
Furthermore, suppose the past wage under training $Y_{0}(1)$ is not informative about the mean of past wage without training $Y_{0}(0)$, once controlling for the initial heterogeneity $U_{0}$.
Then the parallel trends condition holds. \medskip

\ \ 

\noindent \textit{Example 4. (Learning under control.)} As in Example 1, suppose the potential outcome is an individual $i$'s employment status at time $t$, i.e., $Y_{t}(0),Y_{t}(1)\in \{0,1\}$. 
As in Example 2, let the sequence of decisions to join the job training program be made by rational, forward-looking individuals under (DUM).
For simplicity, suppose there is a pre-treatment period, i.e., no one joins the job training program initially so that $D_{0}=0$ for all individuals.
If the past employment status without training $Y_0(0)$ is informative about future employment status without training $Y_1(0)$ for the individual (that is, $Y_0(0)$ affects the distribution of $Y_1(0)$ even after controlling for initial heterogeneity $U_0$), then the parallel trends condition is typically violated --- unless such learning is never valuable (in that it affects the enrollment decision in period 1), or where it is valuable it is also perfect (in that valuable learners become fully aware of their future employment status in period 1 if they were to remain un-enrolled).
Note that the learning channel would invalidate parallel trends even if the treatment decision in period 1 were not a direct function of employment status in period 0, i.e.~recruiting for job training among the unemployed.

\ \ 

\noindent \textit{Example 5. (Selection on present outcomes: static, repeated Roy.)} As in Example 1, suppose the potential outcome is an individual $i$'s employment status at time $t$, i.e., $Y_{t}(0),Y_{t}(1)\in \{0,1\}$. 
Also, as in Example 2, let the sequence of decisions to join job training be made by rational, forward-looking individuals under (DUM).
For simplicity, suppose that opportunity costs of the training program are known and negligible $k=0$, and that individuals know their potential employment status in each period. 
Suppose individuals enroll in job training in period $t$ if it causes employment, $Y_{t} (1) > Y_{t} (0)$.%
\footnote{
    In the original Example 5 we consider a deterministic choice rule that breaks ties in favor of treatment, but the results are essentially unchanged for the deterministic choice rule that breaks ties in favor of no treatment.}
Then the parallel trends condition holds only if there is no aggregate job growth without job training, $E [ Y_{i0} (0) ] = E [ Y_{i1} (0) ]$, and those receiving job training in either period would have otherwise been unemployed in both periods, $Y_{i0} (0) = Y_{i1} (0) = 0 $.

\ \ 

\newpage
\renewcommand*{\bibfont}{\small}
\bibliographystyle{econometrica}
\bibliography{bibliography}

\end{document}